%% file: LieTrafosAndCanonicAngles.tex
\documentclass[
	aps, prd, reprint, a4paper,
	amsmath, amssymb, amsfonts, eqsecnum,
	superscriptaddress,
	showpacs, showkeys,
	nofootinbib, 
]{revtex4-1}

\usepackage{ifthen}
\newboolean{prd}
\setboolean{prd}{true}
\newboolean{arxiv}
\setboolean{arxiv}{false}

\newboolean{notprd}
\setboolean{notprd}{true}
\ifprd
\setboolean{notprd}{false}
\fi
\newboolean{notarxiv}
\setboolean{notarxiv}{true}
\ifarxiv
\setboolean{notarxiv}{false}
\fi

\usepackage{xcolor}
\usepackage{dsfont}
\usepackage{graphicx}

\ifnotprd
\xdefinecolor{mylinkcolor}{rgb}{0,0,0.5}
\usepackage[
	bookmarksnumbered, bookmarksopen, bookmarksopenlevel=2,
\ifnotarxiv
	breaklinks=true,
\fi
	colorlinks=true, filecolor=mylinkcolor, citecolor=mylinkcolor,
	linkcolor=mylinkcolor, urlcolor=mylinkcolor, menucolor=mylinkcolor,
]{hyperref}
\else

\fi

\usepackage[normalem]{ulem}

\newcommand{\okay}{ {} }

\def\vct#1{\mathbf{#1}}

\allowdisplaybreaks

\input{Macros.tex}

\begin{document}
\ifnotprd
\hypersetup{
	pdftitle={Canonical Angles In A Compact Binary Star System With Spinning Components:
	Approximative Solution To Linear Order In Spin For Circular and Conservative Orbits},
	pdfauthor={Manuel Tessmer, Jan Steinhoff, Gerhard Schaefer}
}
\fi

\title{Canonical Angles In A Compact Binary Star System With Spinning Components:
	Approximative Solution Through Next-To-Leading-Order Spin-Orbit Interaction
for Circular Orbits}

\author{Manuel Tessmer}
\email{M.Tessmer@uni-jena.de}
\affiliation{Theoretisch--Physikalisches Institut,
	Friedrich--Schiller--Universit\"at,
	Max--Wien--Platz 1, 07743 Jena, Germany, EU}

\author{Jan Steinhoff}
\email{jan.steinhoff@ist.utl.pt}
\affiliation{Centro Multidisciplinar de Astrof\'isica --- CENTRA, Departamento de F\'isica,
	Instituto Superior T\'ecnico --- IST, Universidade T\'ecnica de Lisboa, 
	Avenida Rovisco Pais 1, 1049-001 Lisboa, Portugal, EU}

\author{Gerhard Sch\"afer}
\email{Gerhard.Schaefer@uni-jena.de}
\affiliation{Theoretisch--Physikalisches Institut,
	Friedrich--Schiller--Universit\"at,
	Max--Wien--Platz 1, 07743 Jena, Germany, EU}

\date{\today}

\begin{abstract}
This publication will deal with an explicit determination of the time
evolution of the spin orientation axes and the evolution of the orbital
phase in the case of circular orbits under next-to-leading order spin-orbit
interactions.
We modify the method of Schneider
and Cui proposed in [``Theoreme \"uber Bewegungsintegrale und ihre Anwendungen in Bahntheorien'',
Verlag der Bayerischen Akademie der Wissenschaften, volume 212, 2005.]
to iteratively remove oscillatory terms in the equations of motion
for different masses that were not present in the case of equal masses.
Our smallness parameter is chosen to be the difference of the symmetric
mass ratio to the value 1/4. 
Before the first Lie transformation, the set of conserved quantities consists of the total angular momentum $\vct{J}$,
the amplitudes of the orbital angular momentum and of the spins, $\Ang, S_1,$ and $S_2$.
In contrary, $\SpinTot := | \vSone + \vStwo |$ is not conserved and we wish to
shift its non-conservation to higher orders of the smallness parameter.

We perform the iterations explicitly
to {first} order, while performing higher orders would mean no
structural difference or harder mathematical difficulties.
To apply this method, we develop a canonical system of spin variables
{reduced by the conservation law of total angular momentum, which is
imposed on the phase space as a constraint.}
The result is an asymptotic series in $\epsilon$ that may be truncated
appropriately considering the physical properties of the regarded system.
\end{abstract}


\maketitle
\tableofcontents
\section{Introduction}

Compact binaries are celestial systems that are likely to possess
spin. Astronomical observations even suggest that accreting
black holes can be spun up to more than 98\% of the maximal (extremal)
spin \cite{McClintock:others:2011}. Further, compact objects are important
sources for gravitational waves, and also important for relativistic astrophysics
\cite{OLeary:Kocsis:Loeb:2009}.
A successful analytical method to deal with compact binaries in general
relativity is the post-Newtonian (PN) approximation, which is applicable when the
distances are large and the velocities small compared to the speed of light $c$.
The PN framework reduces the complicated Einstein equations (nonlinear partial differential
equations) to ordinary differential equations. These PN equations of motion are
often encoded in the form of a Lagrangian potential or a Hamiltonian. The goal
of the present paper is to find approximate solutions to these equations of motion
in the case of circular orbits when the spins of the objects are involved.
The method employed makes crucial use of a phase space structure, so we
employ PN results derived within the ADM (=Arnowitt-Deser-Misner)
canonical formalism of general relativity, whose direct outcome
is a description of the spin motion and the motion of the point mass
with positions $\vct{x}_a$ and momenta $\vct{p}_a$ in terms of a
Hamiltonian. For the ADM approach to the PN approximation in the
presence of spin see \cite{Steinhoff:Schafer:2009:2, Steinhoff:2011}
and references therein.

First attempts on spin in the PN approximation have been done in
\cite{Barker:OConnell:1975, DEath:1975,Barker:OConnell:1979, Thorne:Hartle:1985},
where, e.g., the leading-order (LO) results for spin-orbit and spin(1)-spin(2)
effects have been calculated.
Most important for the present paper are spin-orbit interaction Hamiltonians
up to and including next-to-leading order (NLO), which were derived within
the ADM formalism in \cite{Damour:Jaranowski:Schafer:2008:1, Steinhoff:Schafer:Hergt:2008}.
Corresponding equations of motion were already known before \cite{Tagoshi:Ohashi:Owen:2001,
Faye:Blanchet:Buonanno:2006}, and other approaches succeeded at this order, too
\cite{Porto:2010, Levi:2010, Perrodin:2010}. A generalization of the NLO spin-orbit Hamiltonian
to arbitrary many objects was calculated in \cite{Hartung:Steinhoff:2010}. Even the
next-to-next-to-leading order (NNLO) spin-orbit Hamiltonian was derived
\cite{Hartung:Steinhoff:2011:1} and was recently confirmed
\cite{Marsat:Bohe:Faye:Blanchet:2012, Bohe:Marsat:Faye:Blanchet:2012}. Similar results also exist for
spin(1)-spin(2) interaction. The complete NLO Hamiltonian was calculated in
\cite{Steinhoff:Hergt:Schafer:2008:2} and confirmed by \cite{Porto:Rothstein:2008:1, Levi:2008}
(a partial result is contained in \cite{Porto:Rothstein:2006}, see
\cite{Steinhoff:Hergt:Schafer:2008:2, Porto:Rothstein:2008:1} for a discussion). The NNLO
spin(1)-spin(2) interaction was simultaneously derived in a reduced Hamiltonian
form \cite{Hartung:Steinhoff:2011:2} and in a potential form \cite{Levi:2011} (a comparison
is still missing, see \cite{Hergt:Steinhoff:Schafer:2011} for the emerging difficulties).
For the LO spin(1)-spin(1) interaction, see, e.g., \cite{Poisson:1998}. The extension to NLO
succeeded in the form of a fully reduced Hamiltonian \cite{Steinhoff:Hergt:Schafer:2008:1, Hergt:Schafer:2008, Hergt:Steinhoff:Schafer:2010:1} and in the form of a potential
\cite{Porto:Rothstein:2008:2, *Porto:Rothstein:2008:2:err, Steinhoff:Schafer:2009:1}.
The LO Hamiltonians of cubic and quartic order in spin were derived in
\cite{Hergt:Schafer:2008:2, Hergt:Schafer:2008}
(though crosschecks in \cite{Steinhoff:Puetzfeld:2012} suggest that the quartic
order is not complete yet).

The determination of the far-zone gravitational waves generated by compact binaries requires
the knowledge of certain radiation multipoles. The LO spin-orbit and spin(1)-spin(2)
contributions were derived in \cite{Kidder:1995}. Based on the energy and angular
momentum flux obtained there, dissipative effects on the orbital evolution due to
spin were inferred \cite{Gergely:2000} (see \cite{Gergely:Keresztes:2003} for the
spin(1)-spin(1) level). Corresponding PN equations of motion can also be calculated
directly by solving the Einstein equations in the near-zone \cite{Will:2005,
Wang:Will:2007}, see also \cite{Wang:Steinhoff:Zeng:Schafer:2011, Steinhoff:Wang:2009}
for dissipative (explicitly time-dependent) Hamiltonians. The NLO spin-orbit effects on
the energy flux and the consequences for the evolution of the phase were derived in
\cite{Blanchet:Buonanno:Faye:2006}. Source multipole moments up to quadratic order
in the spins sufficient for the 3PN energy flux \cite{Porto:Ross:Rothstein:2010}
and the 2.5PN radiation field \cite{Porto:Ross:Rothstein:2012} were calculated.
The spin contributions to the gravitational wave form were extended to 2PN order
explicitly in \cite{Buonanno:Faye:Hinderer:2012}. Even some spin-dependent hereditary
contributions at 3PN were derived for circular orbits \cite{Blanchet:Buonanno:Faye:2011}.

There already exist extensive banks of wave forms for circular binaries
without spin, for example to be found in \cite{Ajith:Babak:Chen:others:2008, Ajith:Babak:Chen:others:2008}.
Gravitational waves from eccentric compact binaries have been calculated in
\cite{Memmesheimer:Gopakumar:Schafer:2004} (see also references therein) to 3PN
accuracy in the orbital motion. Recently, analytical gravitational wave form 
expressions in the time-Fourier domain for non-spinning coalescing binaries have been calculated through 2PN
in \cite{Tessmer:Schafer:2010, Tessmer:Schafer:2011}.
As one includes spin, the treatment of the orbital motion gets more complicated,
because in general, spin precession equations for the orbital angular momentum
$\vAng$ have to be taken into account.
For the case that the spins are aligned (up-up, down-down, or up-down configurations),
the binary motion and the GW expressions
are given in \cite{Tessmer:Hartung:Schafer:2012}.
As soon as the spins are not aligned to $\vAng$, the spins start precessing
in a complicated manner such that until now,
PN-exact solutions for the spin motion at leading-order spin-orbit interaction
are known for eccentric orbits only for two cases: (1) the single-spin case,
and (2) the equal-mass case \cite{Konigsdorffer:Gopakumar:2005}.
Very recently, in \cite{Blanchet:Buonanno:Faye:2011} an approximative solution
to the circular-orbit problem through leading-order spin-orbit interaction has been
solved, neglecting specific higher-order terms emanating in the time evolution
of their precession angles as well. We will extend this through next-to-leading order
spin-orbit interaction as they follow from recent developments in the ADM formalism.
Thus, we work with canonical variables throughout this article.

The equations of motion following from the mentioned Hamiltonians are ordinary
differential equations and solving them numerically is straightforward and fast.
However, for the purpose of data analysis this may still be too slow if overlap
integrals with theoretical waveforms for many initial conditions need to be
calculated, e.g., \cite{Damour:Iyer:Sathyaprakash:2001}. In particular the presence of spins drastically
increases the dimension of the parameter space and thus the number of initial
conditions that must be considered. Therefore analytic solutions to the
equations of motion including spin \cite{Konigsdorffer:Gopakumar:2005} are of
great importance. Approximate analytic solutions can also be more accurate than the use of
standard numerical techniques, especially when one evolves the system over several
thousands of orbits. {Such situations are not only relevant for gravitational wave
astronomy, but also for predicting the distribution of recoil (``kick'') velocities
for merging black hole binaries due to the emission of gravitational waves 
\cite{Berti:Kesden:Sperhake:2012}.
(There the evolution of the spin orientations during the long inspiral period plays a crucial role.)}
Analytic solutions are also important for hereditary effects
(e.g., tails), as these depend on the full evolution history. The solutions
derived in the present paper can therefore be useful to extend the spin-dependent
tail effects (for circular orbits) presented in \cite{Blanchet:Buonanno:Faye:2011}
to higher orders. As it seems to be impossible to cover the whole parameter
space analytically, further approximations are necessary, such as expansions
around the equal-mass \cite{Tessmer:2009} or circular-orbit%
\footnote{Manuscript unpublished, under preparation.}
cases.

As long as no radiation reaction effects due to the
emission of gravitational waves are regarded, the total
angular momentum defined as the sum of the individual spins
and the orbital angular momentum $\vAng$ is conserved.
This can play the role of a ``weak'' constraint in the sense of Dirac
\cite{Dirac:1964}, also see \cite{Hanson:Regge:Teitelboim:1976}.
The goal is to find a new set of generalized coordinates that fulfill
standard canonical Poisson brackets. This could be done with the help
of Dirac brackets. But if one is able to find generalized canonical coordinates that
can eliminate the constraints by construction, one is not forced to walk this way.
The hope is that, by reducing the number of dynamical variables with
the help of conservation laws, the problem of finding solutions to
the equations of motion simplifies.

Note that the magnitude of $\vAng$ will not be conserved if
spin(1)-spin(2) interactions are included. We conclude that, for those
terms, the Hamiltonian must depend on the conjugate coordinate to
$\ang$, and the full dynamics is not described by precessions
only.
Anyway, as we include purely spin-orbit terms, the problem simplifies
drastically
and collapses to the dynamics of a sequence of precessions
for a first insight.

The present paper is organized as follows.
After introducing dimensionless quantities and some notation in section \ref{Sec::Dimless},
we present the details of the construction of
a reduced phase space for two objects with spin in Section \ref{Sec::CanonicalVars}.
We find that the three spin angle coordinates used in Ref.~\cite{Tessmer:2009}
are canonical, and their corresponding conjugated momenta are given by the
total spin amplitudes along their rotation axes.
Next, we show details of the Lie transformation algorithm in order to
shift periodic terms to higher orders of the smallness parameter (here: deviation
from the equal-mass case) in a perturbation theory in Section \ref{Sec::LieTransform}.
This method is a modified version of the
``modified Cui method'' \cite{Schneider:Cui:2005, Mai:Schneider:Cui:2008}.
This procedure is applied to the binary spin-orbit Hamiltonian at next-to-leading order
in Section \ref{Sec::LieTrafo_BinaryNLOSO}.
There we also apply the method of shifting perturbative
contributions to the Hamiltonian to even higher orders of the smallness parameter.
A very short review how to combine multiple Lie transformations is provided
in Section \ref{Sec::CombiningLieTrafos}, for those readers who like to go further
than us.
Final conclusions and outlook are given in \ref{Sec::Conclusions}.

\section{Dimensionless Quantities}
\label{Sec::Dimless}
For a binary system it is suitable to work with dimensionless quantities only,
which will be used throughout the paper.
This can be achieved by measuring masses in terms of the reduced mass $\mu$ of
the binary, distances in terms of $G M / c^2$, and time intervals in terms of
$G M / c^3$. Here $M$ is the total mass of the binary, $G$ is the Newton
gravitational constant, and $c$ is the speed of light. All quantities appearing
in the present paper are assumed to be measured in these units from now on and
are therefore dimensionless. The only exception is the dimensionless Kerr
parameter of black holes, which is defined by $\chi_a = c S_a / G m_a^2$ where
$S_a$ is the spin-length and $m_a$ the mass of the $a$-th object. This relation
defines $\chi_a$ also for objects other than black holes, e.g., neutron stars
where this parameter can exceed the value 1.
Astrophysically relevant cases for binary black holes are covered by
$\chi_a \lesssim 1$. We therefore talk of rapidly rotating black holes if
$\chi_a \sim 1$ and of slow rotation if $\chi_a \ll 1 $.

The dimensionless masses of the binary's constituents may be written as
\begin{align}
	m_1 = 1 + \rho \,, \quad
	m_2 = 1 + \rho^{-1} \,,
\end{align}
where $\rho \equiv \frac{m_1}{m_2}$ is the mass ratio. Obviously all expressions
can be written in terms of the mass ratio $\rho$ only instead of the individual
masses $m_1$ and $m_2$. We also make use of the symmetric mass ratio
\begin{equation}
\eta = \frac{m_1 m_2}{(m_1 + m_2)^2} = \frac{\rho}{(1+\rho)^2} \,,
\end{equation}
which takes on the value $\frac{1}{4}$ in the equal-mass case and tends to
zero if one of the masses is much smaller than the other. We further introduce
a parameter $\epsilon$ by
\begin{equation}
  \epsilon^2 := \frac{1}{4} - \eta ,
\end{equation}
which measures the deviation from the equal-mass case. It holds
\begin{equation}
  \rho = \frac{1-2\epsilon}{1+2\epsilon} \,, \quad
  m_1 = \frac{2}{1+2\epsilon} \,, \quad
  m_2 = \frac{2}{1-2\epsilon} \,.
\end{equation}
where we assumed $m_1 \leq m_2$.

The relation between dimensionless spin-lengths $S_a$ and dimensionless Kerr
parameters $\chi_a$ of black holes reads
\begin{equation}\label{spinKerr}
S_1 = \rho^{-1} \chi_1 \,, \quad
S_2 = \rho \chi_2 \,.
\end{equation}
Notice that the dimensionless spins $S_a$ are defined analogous to the $\chi_a$
in some other publications, e.g., in \cite{Tessmer:2009}.

For the sake of simplicity and to avoid introducing a new smallness parameter,
$c$ is regarded to be the book-keeping parameter
for the post-Newtonian approximation and does not have the dimension of speed
in the Hamiltonians for the remainder of this article.

\section{Canonical Variables}
\label{Sec::CanonicalVars}

In this section we introduce the canonical variables, or the phase space,
of the binary system. {This phase space structure is helpful for an
application of Lie-series techniques for solving the equations of motion.
Note that Lie transformations can also be applied to noncanonical
sets of variables, see \cite{Cary:Littlejohn:1983}.}
In order to simplify the problem we reduce the dimension of the
phase space by making use of certain conservation laws. This reduction is
most transparent in a specific basis on phase space, which is derived in
the present section.

\subsection{Center of Mass System\label{sec:COM}}
In this section we illustrate how the number of phase space variables can
be reduced by making use of conservation laws. We discuss this reduction by
looking at the well known transition to the center-of-mass frame. The spins are
neglected for now.

\subsubsection{Reduction}
Consider an action $W$ of the form
\begin{equation}
W = \int dt \, ( \vct{p}_1 \cdot \dot{\vct{z}}_1 + \vct{p}_2 \cdot \dot{\vct{z}}_2 - H ) \,,
\end{equation}
with particle positions $\vct{z}_a$ and canonical momenta $\vct{p}_a$.
The equations of motion follow from a variation of the action
and take on the form of Hamilton's equations, where $H$ plays the role
of the Hamiltonian. (Remember that the variations of $\vct{z}_a$ and
$\vct{p}_a$ are considered as independent.) The Poisson brackets read
\begin{equation}
\{ z_1^i , p_{1j} \} = \delta_{ij} \,, \quad
\{ z_2^i , p_{2j} \} = \delta_{ij} \,,
\end{equation}
all other zero. We call the terms in the action involving time derivatives kinematic terms
in the present paper, as they correspond to the Poisson bracket structure. The
interaction is described by the Hamiltonian.

The dimension of the phase space is 12. This dimension can
be reduced by using conserved quantities. First the total linear momentum
$\vct{P}$ is constant (to the post-Newtonian order considered here). The center-of-mass
system is defined by a vanishing total linear momentum,
\begin{equation}
\vct{P} = \vct{p}_1 + \vct{p}_2 = 0 \,.
\end{equation}
Then the position of the center-of-mass is also constant and may be chosen to be
the coordinate origin. The action now reads
\begin{equation}
W = \int dt \, ( \vct{p} \cdot \dot{\vct{r}} - H ) \,,
\end{equation}
where $\vct{p} = \vct{p}_1 = - \vct{p}_2$ and $\vct{r} = \vct{z}_1 - \vct{z}_2$.
Due to translation invariance the Hamiltonian $H$ depends on $\vct{z}_1$ and
$\vct{z}_2$ solely through the combination  $\vct{r} = \vct{z}_1 - \vct{z}_2$. The phase space
now consists of the six variables $\vct{r}$ and $\vct{p}$ only. The Poisson
brackets can be ``read off'' from the kinematic term, i.e.,
\begin{equation}\label{PBrp}
\{ r^i , p_j \} = \delta_{ij} \,,
\end{equation}
all other zero.

The phase space was reduced from
12 to 6 variables with the help of 6 conserved quantities (total linear momentum
and center-of-mass position).
Formally the conservation laws can be treated as constraints on the
phase space. However, such constraints are already fulfilled by
the equations of motion, i.e., they do not produce additional
constraint forces. Instead of working with an action,
constraints on a phase space can be handled using the Dirac
bracket \cite{Dirac:1964, Hanson:Regge:Teitelboim:1976, Hanson:Regge:1974}. In this sense
(\ref{PBrp}) can be considered as a Dirac bracket.

\subsubsection{Orbital angular momentum}
We can reformulate the phase space by introducing angles for $\vct{r}$ and
(orbital) angular momenta. We introduce a rotation matrix $\rotm_I^{\up i}{}_O^{\up j}$,
\begin{equation}
  \rotm_I^{\up i}{}_O^{\up k} \rotm_I^{\up j}{}_O^{\up k} = \delta_{ij} = \rotm_I^{\up k}{}_O^{\up i} \rotm_I^{\up k}{}_O^j{} \,,
\end{equation}
transforming between the frame $O$, which co-rotates with the
orbital motion of the binary, and the inertial frame $I$. Notice that $\rotm_I^{\up i}{}_O^{\up j}$ can be parametrized
by three angle variables, e.g., the Euler angles. We can choose the 1-axis of
the co-rotating frame $\rotm_I^{\up i}{}_O^{\up 1}$ such that it points in the direction of $\vct{r}$,
\begin{equation}\label{Ocond}
  \rotm_I^{\up i}{}_O^{\up 1} = n^i \,,
\end{equation}
where $\vct{n} = \vct{r} / r$ and $r = |\vct{r}|$. The kinematic terms then turn into
\begin{align}
  \vct{p} \cdot \dot{\vct{r}} &= (\vct{n} \cdot \vct{p}) \dot{r}
  + p^i \dot{\rotm}_I^{\up i}{}_O^{\up j} \rotm_I^{\up k}{}_O^{\up j} \rotm_I^{\up k}{}_O^{\up 1} r \,, \\
  &= p_r \, \dot{r} + \frac{1}{2} L_{ki} \avel_{IO}^{ki} \,,
\end{align}
where
\begin{equation}
  p_r := \vct{n} \cdot \vct{p} , \quad
  L_{ki} := 2 r^{[k} p^{i]} , \quad
  \avel_{IO}^{ki} := \rotm_I^{\up k}{}_O^{\up j} \dot{\rotm}_I^{\up i}{}_O^{\up j} \,.
\end{equation}
Notice that the angular velocity tensor $\avel_{IO}^{ki}$ is antisymmetric,
$\avel_{IO}^{ki} = - \avel_{IO}^{ik}$. $L_{ki}$ is the orbital angular momentum tensor. Corresponding
vectors are given by
\begin{equation}
L^i := \frac{1}{2} \epsilon_{ijk} L_{jk} , \quad
\avel_{IO}^i := \frac{1}{2} \epsilon_{ijk} \avel_{IO}^{jk} \,.
\end{equation}
The implications of this new form of the kinematic terms for the phase
space structure are discussed in the following.

\subsection{Canonical angles: general theory}
In the center-of-mass frame, most of the dynamics of a binary system
with spinning components can be described in terms of angular momenta,
namely the orbital angular momentum and the two spins of the components.
Before we derive canonical variables for this system, let us
prepare some general developments on the theory of classical angular
momenta.

\subsubsection{Angular momentum algebra}
We consider a generic angular momentum represented by an antisymmetric
tensor $S_{ij} = - S_{ji}$ and a rotation matrix $\rotm_I^{\up i}{}_C^{\up j}$ transforming to
a ``co-rotating frame $C$. Inspired by the kinematic terms found in the
last section, we consider an action of the form
\begin{equation}
W = \int dt \, \left[ \frac{1}{2} S_{ij} \avel_{IC}^{ij} - H( \rotm_I^{\up i}{}_C^{\up j}, S_{ij} ) \right] \,.
\end{equation}
Let us derive the equations of motion by independently varying
$S_{ij}$ and the angle variables parameterizing $\rotm_I^{\up i}{}_C^{\up j}$.
The details of this process are analogous to the relativistic
case discussed in \cite{Hanson:Regge:1974} and are not repeated here.
(For example, it is easiest to use the independent antisymmetric
variation symbol $\delta \theta^{ij} = - \delta \theta^{ji} = \rotm_I^{\up i}{}_C^{\up k} \delta \rotm_I^{\up j}{}_C^{\up k}$ instead of explicitly varying the angle variables.)
Again, the equations of motion can be written as Hamilton's
equations (with Hamiltonian $H$) if we impose the Poisson
brackets
\begin{gather}
  \{ \rotm_I^{\up i}{}_C^{\up j}, \rotm_I^{\up k}{}_C^{\up l} \} = 0 \,, \quad \{ \rotm_I^{\up i}{}_C^{\up j}, S_{kl} \} = \delta_{il} \rotm_I^{\up k}{}_C^{\up j} - \delta_{ik} \rotm_I^{\up l}{}_C^{\up j} \,, \label{nrPB1} \\
\{ S_{ij}, S_{kl} \} = \delta_{ik} S_{jl} - \delta_{jk} S_{il}
	- \delta_{il} S_{jk} + \delta_{jl} S_{ik} \,. \label{nrPB2}
\end{gather}
Notice that the last relation is the usual angular momentum algebra,
\begin{equation}
  \{ S^i, S^j \} = \epsilon_{ijk} S^k \,,
\end{equation}
where $S^i := \frac{1}{2} \epsilon_{ijk} S_{jk}$ is the spin vector.
If the Hamiltonian $H$ does not depend on the orientation of the
co-rotating frame, or $\rotm_I^{\up i}{}_C^{\up j}$, then only this angular momentum algebra is
needed. However, for now we look at the most general case where
the phase space consists of three pairs of canonical variables
contained in $\rotm_I^{\up i}{}_C^{\up j}$ and $S_{ij}$.

\subsubsection{Euler angles}
Before we can make the phase space structure more explicit, we
need a way to parametrize a generic rotation matrix. One possibility
is
\begin{equation}
\rotm^{ij}( \omega , \vct{n} ) = n^i n^j + ( \delta_{ij} - n^i n^j ) \cos\omega - \epsilon_{ijk} n^k \sin\omega \,,
\end{equation}
which gives a rotation of angle $\omega$ around a unit vector $\vct{n}$,
which intrinsically provides the rotation axis. However, throughout
this paper we will parametrize rotation matrices in terms of three
Euler angles $\alpha$, $\beta$, and $\gamma$ by
\begin{equation}
  \rotm^{ij}( \alpha, \beta, \gamma ) = \rotm^{ik}( \alpha, \vct{e}_3 ) \rotm^{kl}( \beta, \vct{e}_1 ) \rotm^{lj}( \gamma, \vct{e}_3 ) \,,
\end{equation}
where $\vct{e}_a = ( \delta_a^i )$. Notice that it holds $\rotm^{ji}( \alpha, \beta, \gamma ) = \rotm^{ij}( - \gamma, - \beta, - \alpha )$.
The Euler angles always enter the kinematic terms through
the angular velocity
\begin{equation}
\avel^i( \alpha, \beta, \gamma ) = \frac{1}{2} \epsilon_{ijk} \rotm^{jl}( \alpha, \beta, \gamma ) \dot{\rotm}^{kl}( \alpha, \beta, \gamma ) \,.
\end{equation}
A straightforward calculation leads to
\begin{equation}\label{euler_vel}
\left( \avel^i(\alpha, \beta, \gamma) \right) = \left( \begin{array}{c}
	\cos\alpha  \, \dot{\beta} + \sin\alpha \sin\beta \, \dot{\gamma} \\
	\sin\alpha \, \dot{\beta} - \cos\alpha \sin\beta \, \dot{\gamma} \\
	\dot{\alpha} + \cos\beta \, \dot{\gamma}
\end{array} \right) \,.
\end{equation}
This result will be used frequently throughout the next
sections.

\subsubsection{Canonical angle variables\label{spincan}}
Let us introduce another frame $S$ in which the 3-axis is aligned
to $S_i$ (which in general is not
necessarily identical to the co-rotating frame, e.g., for a
non-spherical top).
We parametrize the transformation as
\begin{equation}
  \rotm_I^{\up i}{}_S^{\up j} = \rotm^{ij}( \phi_S, \theta_S, 0 ) \,,
\end{equation}
such that
\begin{equation}
( S^i ) = S ( \rotm_I^{\up i}{}_S^{\up 3} ) = S \left( \begin{array}{c}
	\sin\theta_S \sin\phi_S \\
	- \sin\theta_S \cos\phi_S \\
	\cos\theta_S
\end{array} \right) \,,
\end{equation}
where $S := |\vct{S}|$. We can then decompose the transformation to
the co-rotating frame C as $\rotm_I^{\up i}{}_C^{\up j} = \rotm_I^{\up i}{}_S^{\up k} \rotm_S^{\up k}{}_C^{\up j}$ and parametrize $\rotm_S^{\up i}{}_C^{\up j}$
in terms of Euler angles,
\begin{equation}
  \rotm_S^{\up i}{}_C^{\up j} = \rotm^{ij}(\alpha_S, \beta_S, \gamma_S) \,.
\end{equation}
However, as the first and the last of the Euler angles is a
rotation around the 3-axis, it holds
\begin{multline}
  \rotm^{ik}( \phi_S, \theta_S, 0 ) \rotm^{kj}(\alpha_S, \beta_S, \gamma_S) \\
  = \rotm^{ik}( \phi_S, \theta_S, \alpha_S ) \rotm^{kj}(0 , \beta_S, \gamma_S) .
\end{multline}
Thus an equivalent formulation is
\begin{equation}\label{newsplit}
  \rotm_I^{\up i}{}_S^{\up j} = \rotm^{ij}( \phi_S, \theta_S, \alpha_S ) \,, \quad
  \rotm_S^{\up i}{}_C^{\up j} = \rotm^{ij}( 0, \beta_S, \gamma_S ) \,,
\end{equation}
which in hindsight of later calculations is used from now
on. The reason is that if the rotation axis of the
co-rotating frame is parallel to the spin, then $\beta_S$ and
$\gamma_S$ are constant and can usually even be set to zero.
Then with the convention (\ref{newsplit}) the frames $S$ and
$C$ coincide.

Using the decomposition
$\rotm_S^{\up i}{}_C^{\up j} = \rotm_I^{\up k}{}_S^{\up i} \rotm_I^{\up k}{}_C^{\up j}$ we obtain an
``addition theorem'' for the angular velocity,
\begin{equation}
  \avel_{SC}^{ij} = \avel_{IC}^{kl} \rotm_I^{\up k}{}_S^{\up i} \rotm_I^l{}_S^j + \avel_{SI}^{ij} \,,
\end{equation}
with the angular velocity of the frame $C$ relative to the frame $S$,
$\avel_{SC}^{ij} := \rotm_S^{\up i}{}_C^{\up k} \dot{\rotm}_S^{\up j}{}_C^{\up k}$,
and the angular velocity of the inertial frame $I$ relative to the frame $S$,
$\avel_{SI}^{ij} := \rotm_S^{\up i}{}_I^{\up k} \dot{\rotm}_S^{\up j}{}_I^{\up k}$ (where $ \rotm_S^{\up i}{}_I^{\up j} =  \rotm_I^{\up j}{}_S^{\up i}$).
In terms of angular velocity vectors this reads
\begin{equation}\label{avel_add}
  \avel_{SC}^i = \rotm_I^{\up j}{}_S^{\up i} \avel_{IC}^j + \avel_{SI}^i \,,
\end{equation}
where we used $ \epsilon_{ijk} \rotm_I^{\up l}{}_S^{\up i} \rotm_I^{\up m}{}_S^{\up j} \rotm_I^{\up n}{}_S^{\up k} = \epsilon_{lmn}$.
We can now rewrite the kinematic terms in the frame $S$,
\begin{align}
  \frac{1}{2} S_{ij} \avel_{IC}^{ij} &= S^i \avel_{IC}^{i}
  = S \rotm_I^{\up i}{}_S^{\up 3} \avel_{IC}^i \,, \\
  &= S \avel_{SC}^3 - S \avel_{SI}^3 \,.
\end{align}
Using
\begin{equation}
  \avel_{SI}^{i} = \Omega^i(-\alpha_S, -\theta_S, -\phi_S) \,, \quad
  \avel_{SC}^{i} = \Omega^i(0, \beta_S, \gamma_S) \,,
\end{equation}
and (\ref{euler_vel}) we finally have
\begin{align}
  \frac{1}{2} S_{ij} \avel_{IC}^{ij} &=
  S \cos\beta_S \dot{\gamma}_S + S \dot{\alpha}_S + S \cos\theta_S \dot{\phi}_S \,, \\
  &= S^3 \dot{\phi}_S + S^3_C \dot{\gamma}_S + S \dot{\alpha}_S \,,
\end{align}
where $S^3 := S \cos\theta_S$ is the 3-component of the spin vector in the inertial frame
and $S^3_C := S \cos\beta_S$ is the 3-component of the spin vector in the co-rotating frame.
The Poisson brackets read
\begin{equation}\label{PBonespin}
  1 = \{ \phi_S, S^3 \} = \{ \gamma_S, S^3_C \} = \{ \alpha_S, S \} \,,
\end{equation}
all other zero. We have explicitly expressed $\rotm_I^{\up i}{}_C^{\up j} = \rotm_I^{\up i}{}_S^{\up k} \rotm_S^{\up k}{}_C^{\up j}$
and $S_{ij}$ in terms of three pairs of canonical variables.

In the case that the Hamiltonian $H$ is independent of the orientation
of the co-rotating frame, that is, of $\gamma_S$, $S^3_C$, and $\alpha_S$,
the phase space can be further simplified. The canonical conjugates
$S^3_C$, $\gamma_S$, and $S$ are then cyclic and therefore constant.
The only relevant Poisson bracket is $\{ \phi_S, S^3 \} = 1$ in this case.

\subsubsection{Example: Orbital angular momentum}
As an example we apply our findings from the last section to the orbital
angular momentum. We parametrize the frame $L$, in which the orbital
angular momentum is parallel to the 3-direction, as
\begin{equation}
  \rotm_I^{\up i}{}_L^{\up j} = \rotm^{ij}( \phi_L, \theta_L, \alpha_L ) \,,
\end{equation}
such that
\begin{equation}
( L^i ) = L ( \rotm_I^{\up i}{}_L^{\up 3} ) = L \left( \begin{array}{c}
	\sin\theta_L \sin\phi_L \\
	- \sin\theta_L \cos\phi_L \\
	\cos\theta_L
\end{array} \right) \,.
\end{equation}
As $\vct{L}$ is orthogonal to $\vct{n}$, we can choose the angle $\alpha_L$ such
that $\vct{n}$ is aligned with the 1-axis in the $L$ frame. Then
the frame co-rotating with the orbit $O$ can be chosen to be identical to the
frame $L$, $\rotm_L^{\up i}{}_O^{\up j} = \rotm^{ij}(0, 0, 0)$, as this
choice satisfies the condition (\ref{Ocond}).
According to the last section it holds
\begin{equation}
\frac{1}{2} L_{ki} \avel_{IO}^{ki} = L_{i} \avel_{IL}^i = L^3 \dot{\phi}_L + L \dot{\alpha}_L \,,
\end{equation}
and the Poisson brackets read
\begin{equation}\label{PBorbit}
  1 = \{ \phi_L, L^3 \} = \{ \alpha_L, L \} \,,
\end{equation}
all other zero.

\subsubsection{Addition of angular momenta\label{angle_add}}
We now consider the case that we have two spins $\vct{S}_1$ and
$\vct{S}_2$ which add up to a total spin $\vct{S} := \vct{S}_1 + \vct{S}_2$.
Of course, we could just construct the phase space by
two copies of (\ref{PBonespin}). However, we try
to include the total spin $\vct{S}$ in the phase space here.

We utilize frames denoted by $S_1$, $S_2$, and $S$ where the corresponding
angular momenta point into the 3-direction, respectively.
As the vectors $\vct{S}_1$, $\vct{S}_2$, and $-\vct{S}$
form a triangle and thus lie in a plane, it makes sense
to introduce a frame in which this plane is fixed as, say,
the 2-3-plane. This can be achieved by rotating the frame $S$
around $\vct{S}$ by a suitable angle $\alpha_S$, so we again have
\begin{equation}
  \rotm_I^{\up i}{}_S^{\up j} = \rotm^{ij}( \phi_S, \theta_S, \alpha_S ) \,.
\end{equation}
As the spins $S_a$ lie in the 2-3-plane within the frame
S, we can write
\begin{equation}
  \rotm_S^{\up i}{}_{S_a}^{\up j} = \rotm^{ij}( 0, \theta_a, \alpha_a ) \,,
\end{equation}
where $a=1,2$. Notice that the angles $\theta_a$ are fixed,
as the length of the edges of the triangle are given by
$S_1 := |\vct{S}_1|$, $S_2 := |\vct{S}_2|$, and $S := |\vct{S}|$.
We further introduce co-rotating frames
$C_a$ for the spins $S_a$. The remaining transition to the
co-rotating frames is parametrized as
\begin{equation}
  \rotm_{S_a}^{\up i}{}_{C_a}^{\up j} = \rotm^{ij}( 0, \beta_a, \gamma_a ) \,.  
\end{equation}

The decomposition $\rotm_I^{\up i}{}_{C_a}^{\up j} = \rotm_I^{\up i}{}_S^{\up k} \rotm_S^{\up k}{}_{C_a}^{\up j}$ leads to
\begin{equation}
  \avel_{I C_a}^i = \rotm_I^{\up i}{}_S^{\up j} \avel_{SC_a}^j + \avel_{IS}^i \,,
\end{equation}
cf.~(\ref{avel_add}). Then the kinematic terms can be
written as
\begin{equation}
  \sum_a \frac{1}{2} S_{a ij} \avel_{IC_a}^{ij} 
  = S^i \avel_{IS}^i + \sum_a S^i_a \rotm_I^{\up i}{}_S^{\up j} \avel_{SC_a}^j \,.
\end{equation}
These terms can be evaluated as in Sec.\ \ref{spincan}. For
the first term it immediately follows that
\begin{equation}
  S^i \avel_{IS}^i = S \dot{\alpha}_S + S \cos\theta_S \dot{\phi}_S \,.
\end{equation}
In the other terms we insert
\begin{align}
  S_a^i &= S_a \rotm_I^{\up i}{}_S^{\up j} \rotm_S^{\up j}{}_{S_a}^{\up 3} \,, \\
  \rotm_S^{\up j}{}_{S_a}^{\up i} \avel_{S C_a}^j &= \avel_{S_a C_a}^i - \avel_{S_a S}^i \,,
\end{align}
and finally obtain
\begin{equation}
  \sum_a \frac{1}{2} S_{a ij} \avel_{IC_a}^{ij} =
  S^3 \dot{\phi}_S + S \dot{\alpha}_S
  + \sum_a ( S^3_{a C} \dot{\gamma}_a + S_a \dot{\alpha}_a ) \,,
\end{equation}
where $S^3 := S \cos\theta_S$ and $S^3_{aC} := S_a \cos\beta_a$.
The Poisson brackets read
\begin{equation}
  1 = \{ \phi_S, S^3 \} = \{ \alpha_S, S \}
  = \{ \gamma_a, S^3_{aC} \} = \{ \alpha_a, S_a \} \,,
\end{equation}
all other zero. The dimension of the phase space is 12,
as expected.

Remember that the angles $\theta_a$ are not part of the
phase space, but must be fixed from geometrical
considerations in terms of the other variables.
As the vectors $\vct{S}_1$, $\vct{S}_2$, and $-\vct{S}$
form a triangle and its lengths are part of the
phase space, the angles $\theta_a$ can be
obtained from the law of Cosines (see also
Sec.\ \ref{redux}).

\subsection{Canonical angles: binary system}
\label{SubSec::CanonicalAngles_BinarySystem}
We are now going to construct the phase space of a binary
system with spinning components. We have just seen how
the spins $\vct{S_1}$ and $\vct{S}_2$ of the components
combine to the total spin $\vct{S} := \vct{S}_1 + \vct{S}_2$. 
Now we also add the orbital angular momentum $\vct{L}$ to form the
total angular momentum $\vct{J}$ of the binary system,
\begin{equation}
\vct{J} = \vct{L} + \vct{S} \,.
\end{equation}
Notice that $\vct{J}$ is conserved to the post-Newtonian order
considered here, which we use to reduce the number of variables.

Notice that one could build the phase space simply
by (\ref{PBorbit}) and two copies of (\ref{PBonespin}). However, as we want
to make use of the conservation of $\vct{J}$ later, it is
convenient to include $\vct{J}$ in the phase space.

\subsubsection{Complete phase space}
In Sec.\ \ref{angle_add} we found that
\begin{equation}
  \sum_a \frac{1}{2} S_{a ij} \avel_{IC_a}^{ij} 
  = S^i \avel_{IS}^i + \sum_a ( S^3_{a C} \dot{\gamma}_a + S_a \dot{\alpha}_a ) \,.
\end{equation}
We will now apply the same procedure to the terms
\begin{equation}
  L^{i} \avel_{IL}^{i} + S^i \avel_{IS}^i \,.
\end{equation}
That is, we introduce a frame $J$ by
\begin{equation}
  \rotm_I^{\up i}{}_J^{\up j} = \rotm^{ij}( \phi_J, \theta_J, \alpha_J ) \,,
\end{equation}
such that $J^i = J \rotm_I^{\up i}{}_J^{\up 3} $ (here $J := |\vct{J}|$)
and the vectors $\vct{L}$ and $\vct{S}$ are lying in the
2-3-plane in the frame $J$. The parameterization of
the frames $L$ and $S$ is now given relative to the
frame $J$ as
\begin{equation}
  \rotm_J^{\up i}{}_L^{\up j} = \rotm^{ij}( 0, \theta_L, \alpha_L ) \,, \quad
  \rotm_J^{\up i}{}_S^{\up j} = \rotm^{ij}( 0, \theta_S, \alpha_S ) \,.
\end{equation}
The result from Sec.\ \ref{angle_add} translates into
\begin{equation}
  L^{i} \avel_{IL}^{i} + S^i \avel_{IS}^i
  = J^3 \dot{\phi}_J + J \dot{\alpha}_J + L \dot{\alpha}_L + S \dot{\alpha}_S \,,
\end{equation}
where $J^3 := J \cos\theta_J$. The angles $\theta_L$ and $\theta_S$
must be obtained from geometric considerations.

In total we have
\begin{multline}
  p_i \dot{r}^i + \sum_a \frac{1}{2} S_{a ij} \avel_{IC_a}^{ij}
  = p_r \dot{r} + J^3 \dot{\phi}_J + S^3_{1 C} \dot{\gamma}_1 + S^3_{2 C} \dot{\gamma}_2 \\
  + J \dot{\alpha}_J + L \dot{\alpha}_L + S \dot{\alpha}_S
  + S_1 \dot{\alpha}_1 + S_2 \dot{\alpha}_2 \,,
\end{multline}
and the Poisson brackets read
\begin{equation}
\begin{split}
  1 &= \{ r, p_r \} = \{ \phi_J, J^3 \}
  = \{ \gamma_1, S^3_{1C} \} = \{ \gamma_2, S^3_{2C} \} \\
  &= \{ \alpha_J, J \} = \{ \alpha_L, L \} = \{ \alpha_S, S \} \\
  &= \{ \alpha_1, S_1 \} = \{ \alpha_2, S_2 \} \,,
\end{split}
\end{equation}
all other zero.

\subsubsection{Reduced phase space\label{redux}}
As in Sec.\ \ref{sec:COM} we are now going to utilize
conserved quantities to reduce the number of phase space
variables.

We assume that the Hamiltonian is independent
of the orientation of the co-rotating frames of the two
components. That is, the Hamiltonian is independent of
$\gamma_a$, $S^3_{aC}$, and $\alpha_a$. Then the canonical conjugates $S^3_{aC}$,
$\gamma_a$, and $S_a$ are constant. The corresponding kinematic
terms therefore turn into total time derivatives and can be
dropped.

Next we make use of the conservation of total angular
momentum $\vct{J}$. This allows one to align $\vct{J}$
with the 3-axis, i.e.,
\begin{equation}
  J^3 = J \, \quad \Leftrightarrow \quad \theta_J = 0 \,.
\end{equation}
In order to make contact
with the notation in \cite{Tessmer:2009}, introduce
the alternative notation
\begin{gather}
  \alpha_J + \phi_J \equiv \Upsilon \,, \quad
  \alpha_L \equiv \varphi \,, \quad
  \alpha_S \equiv \phi_S - \frac{\pi}{2} \,, \\
  \theta_L \equiv \Theta \,, \quad
  \theta_S = \Theta - \alpha_{\rm ks} \,, \quad
  \theta_1 \equiv \tilde{s} \,, \quad
  \theta_2 = \tilde{s} + \alpha_{12} - \pi \,.
\end{gather}
The kinematic terms simplify to
\begin{multline}
  p_i \dot{r}^i + \sum_a \frac{1}{2} S_{a ij} \avel_{IS_a}^{ij}
  = p_r \dot{r} + J \dot{\Upsilon}
  + L \dot{\varphi} + S \dot{\phi}_S \,,
\end{multline}
and the Poisson brackets are
\begin{equation}
  1 = \{ r, p_r \} = \{ \Upsilon, J \} = \{ \varphi, \Ang \} = \{ \phi_S, \SpinTot \} \,,
\end{equation}
all other zero. There is still one conserved quantity left,
namely $J$. This means that the Hamiltonian is independent
of $\Upsilon$, and one could even drop $J$ and $\Upsilon$
from the phase space, too. However, $\Upsilon$ is needed
for gravitational wave forms, so we will keep it for now.
(Similarly, if one could observe the absolute orientation
of the components somehow, then one would like to keep
the variables $\alpha_1$ and $\alpha_2$, too.)

Notice that beforehand we have formulated the phase space in
a way that optimally allowed us to implement the conservation
laws as constraints. This corresponds to the change from 
Lagrange equations of the first kind to
Lagrange equations of the second kind
in such a way that we have found generalized coordinates that
eliminate constraints on the dynamical variables.

\begin{figure}[ht!]
  \includegraphics[width=\textwidth, angle=-0,scale=0.4]{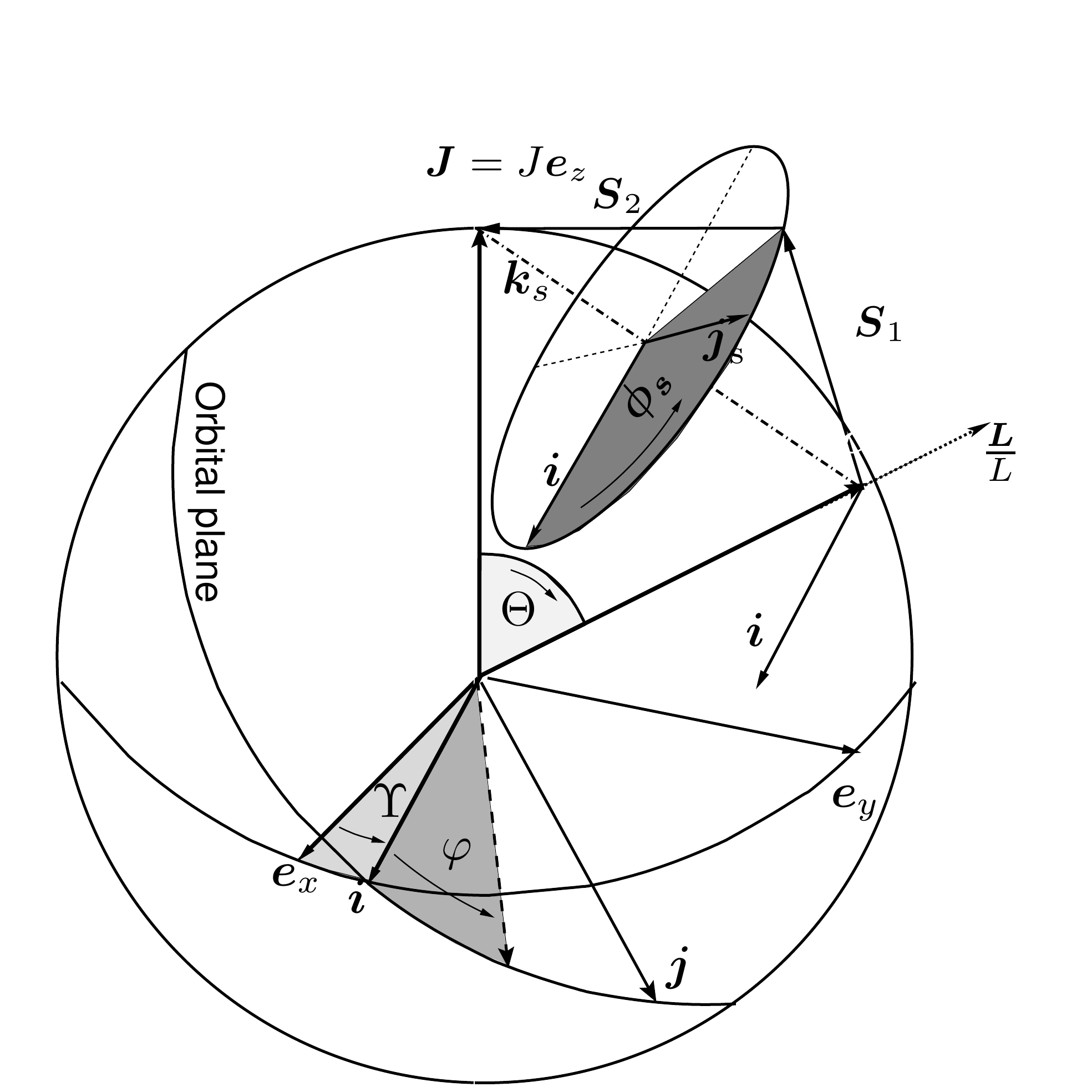}
  \caption{The geometry of the rotation angles. 
The dashed line represents the vector from the center of mass to the reduced mass.
The vectors ($\vct{i}, \vct{j}$) span the orbital plane perpendicular to
$\vct{k}_\ang \defdby \frac{\vAng}{\Ang}$.
The vectors ($\vct{i}, \vct{j}_S$) span the plane perpendicular to
$\vct{k}_S \defdby \frac{\vct{\SpinTot}}{\SpinTot}$.
\label{Fig:FigSpinorbitLie}}
\end{figure}

Let us summarize the transition to the new variables, which
must be inserted into the Hamiltonian:
\begin{align}
  r^i &= r \rotm^{i1}( \Upsilon, \Theta, \varphi ) \,, \\
  p_i &= p_r \rotm^{i1}( \Upsilon, \Theta, \varphi )  + \frac{L}{r} \rotm^{i2}( \Upsilon, \Theta, \varphi ) \,, \\
  S_1^i &= S_1 \rotm^{ij}( \Upsilon, \theta_S, \phi_S - \pi/2) \rotm^{j3}( 0, \theta_1, 0 ) \,, \\
  S_2^i &= S_2 \rotm^{ij}( \Upsilon, \theta_S, \phi_S - \pi/2) \rotm^{j3}( 0, \theta_2, 0 ) \,.
\end{align}
Remember that we need to solve for $\Theta$, $\theta_S$,
$\theta_1$, and $\theta_2$, using geometric considerations (we know the length
of all edges in the triangle where they appear, so one can
apply the law of Cosines and Sines),
\begin{align}
  \cos \Theta &= \frac{J^2 + L^2 - S^2}{2 J L} \,, \\
  \cos \theta_S &= \frac{J^2 + S^2 - L^2}{2 J S} \,, \\
  \cos \theta_1 &= \frac{S^2 + S_1^2 - S_2^2}{2 S S_1} \,, \\
  \cos \theta_2 &= \frac{S^2 + S_2^2 - S_1^2}{2 S S_2} \,, \\
  S_1 \sin \theta_1 &= - S_2 \sin \theta_2 \,, \\
  L \sin \Theta &= - S \sin \theta_S \,.
\end{align}
Notice that from now on we are not utilizing upper indices to
denote vector components any more, upper indices are always
exponents. Alternatively for $\tilde{s}$, $\alpha_{\rm ks}$
and $\alpha_{12}$ we have, see \cite{Tessmer:2009} and
Fig.~\ref{Fig:FigSpinorbitLie},
\begin{align}
\tilde{s} &= \sin ^{-1}\left(
\frac{  S_2}{\SpinTot}
	\sqrt{1-\frac{\left(S_1^2+S_2^2-\SpinTot^2\right)^2}{4 S_1^2 S_2^2}}\right) \,, \\
\alpha_{12} &= \cos ^{-1}\left(\frac{\SpinTot^2-S_1^2-S_2^2}{- 2 S_1 S_2}\right) \,, \\
\alpha_{\rm ks} &=
\pi -\sin ^{-1}\left(\frac{J}{\SpinTot} 
  \sqrt{1-\frac{\left(J^2+L^2-\SpinTot^2\right)^2}{4 J^2 L^2}}\right) \,.
\end{align}
However, we will actually \emph{not utilize} these inverse trigonometric
functions, as they only give unique values if additional assumptions
on the spin orientations are made. Still Poisson brackets can be uniquely
evaluated as follows. Partial derivatives of the constrained angles
with respect to the canonical variables can be obtained by differentiating
the law of cosines, e.g.,
\begin{equation}
  - \sin \Theta(J, L, S) \frac{\partial \Theta(J, L, S)}{\partial S} = - \frac{S}{J L}
\end{equation}
We actually substitute the Cosines of the constrained angles
in terms of canonical variables using the law of Cosines. The Sines are
first reduced to $\sin \alpha_{\text{ks}}$ and $\sin \alpha_{12}$ using
\begin{align}
  \frac{\sin \Theta}{S} = -\frac{\sin \theta_S}{L}
  &= \frac{\sin \alpha_{\text{ks}}}{J} \,, \\
  \frac{\sin \theta_1}{S_2} = - \frac{\sin \theta_2}{S_1} &= \frac{\sin \alpha_{12}}{S} \,.
\end{align}
Then it holds
\begin{align}
  \frac{\partial \sin \alpha_{12}(S, S_1, S_2)}{\partial S}
  &= \frac{S (S_1^2 + S_2^2 - S^2)}{2 S_1^2 S_2^2 \sin \alpha_{12}(S, S_1, S_2)} \,, \label{Dalpha1} \\
  \frac{\partial \sin \alpha_{\text{ks}}(J, L, S)}{\partial J}
  &= \frac{J (L^2 + S^2 - J^2)}{2 L^2 S^2 \sin \alpha_{\text{ks}}(J, L, S)} \,, \label{Dalpha2} \\
  \frac{\partial \sin \alpha_{\text{ks}}(J, L, S)}{\partial L}
  &= \frac{(J^2 - S^2)^2 - L^2}{4 L^3 S^2 \sin \alpha_{\text{ks}}(J, L, S)} \,, \label{Dalpha3} \\
  \frac{\partial \sin \alpha_{\text{ks}}(J, L, S)}{\partial S}
  &= \frac{(J^2 - L^2)^2 - S^2}{4 L^2 S^3 \sin \alpha_{\text{ks}}(J, L, S)} \,. \label{Dalpha4}
\end{align}
These relations are used extensively throughout the present publication
for the calculation of Poisson brackets.

\section{The Method For Eliminating Periodic Terms in Canonical EOM}
\label{Sec::LieTransform}

In this section, we summarize some main ideas how to eliminate
purely periodic perturbation terms in Hamiltonian functions
explained in \cite{Schneider:Cui:2005,Mai:Schneider:Cui:2008}.
Several variants of this method are discussed in the review
\cite{Cary:1981}.
The linchpin there is to apply finite Lie transformations
with a special choice of the generating function.


Let the set $X_a$ represent the canonical coordinates and momenta,
say $x_a=X_a$ and $p_a = X_{n+a}$ for $a=1...n$.
Having chosen a generating function $s$,
the {\em transformed} quantity $X$ will follow as
\begin{eqnarray}
 \hat{X}
	&=& X + \PB{X}{s} + \frac{1}{2!} \PB{\PB{X}{s}}{s} + ... \neanl
	&=& X + {\mathcal D}_s X + \frac{1}{2!}{\mathcal D}_s^2 X + ... \neanl
	&=& X + \sum_{k\ge 1}^{} \frac{1}{k!} {\mathcal D}_s^k X 
	=: e^{{\mathcal D}_s} X \,,
\end{eqnarray}
with
\begin{eqnarray}
 {\mathcal D}_s^0 X &:=& X			\,, \\
 {\mathcal D}_s^1 X & =& \PB{X}{s}		\,, \\
 {\mathcal D}_s^2 X & =& \PB{\PB{X}{s}}{s}	\,,
\end{eqnarray}
and so on. The Poisson brackets $\PB{X_a}{X_{n+a}}$ turn out to be invariant
under this transformation, see, e.g., \cite{Deprit:1969, Schneider:Cui:2005}.
It is therefore just a particular canonical transformation.

To make this transformation valuable, we let the Hamiltonian function consist
of an integrable part $F^*$ and a term ${R}^{(1)}$ which is purely oscillatory
(as the solution to $F^*$ is inserted),
\begin{equation}
 \HAM{}{}(X) = F^{*}(X) + {R}^{(1)}(X) \,.
\end{equation}
with $\order{R^{(1)}}{}=\epsilon$ and $\epsilon \ll 1$ as well.
For the dynamics of $F^{*}$ alone, an analytical solution to the system of
canonical equations of motion may be known, $X_{a, {\rm solution}}~=:~\bar{X}_a(t)$.
Then, via a contact transformation, we are able to shift the
perturbation term to order $\order{\epsilon}{2}$. This works
well if we choose the generating function to be
\begin{equation}
\label{Eq::DefGenerator}
 s = \int_{t_0}^{t} {R}|_{X=\bar{X}(t')} \, {\rm d} t'
\end{equation}
and re-express the result purely in terms of canonical variables
again\footnote{One could also use a Fourier decomposition of the
residue and the generating function and obtain an algebraic relation
for the coefficients of the latter. In the case of more dimensions
one will be confronted with the problem of small denominators
already for the first order coefficients. The famous KAM theory
overcame this problem which plagued celestial mechanics in the 20$^{\text{th}}$
century.}.
That is, the explicit time dependence is removed
with the help of the solution $\bar{X}(t')=X$ after the integration
was performed with the help of inverting the relation
\begin{equation}
 X = X (t+c_1, c_2, ..., c_{2n}) \,,
\end{equation}
with the $c's$ as a set of 2n integration constants, to finally get
\begin{eqnarray}
t+c_1 &=& \varphi_1 (X) \,, \\
  c_j &=& \varphi_j (X) \,, (j=2...2n) \,.
\end{eqnarray}
From Eq. (\ref{Eq::DefGenerator}) one obtains
\begin{equation}
\PB{F^*}{s}+R=0\,.
\end{equation}
Note that (\ref{Eq::DefGenerator}) holds as we employ
a canonical supplementary system where only the integrable part
of the Hamiltonian
fixes the evolution. After re-expressing each component
of the generator $s$ in terms of canonical coordinates only, it
does not matter using which dynamics $s$ has been computed,
- it has the nice feature
to definitely eliminate the leading-order (in $\epsilon$) terms
of the perturbing Hamiltonian, see the computation below%
\footnote{It is a vital detail that $R$
is assumed to be oscillatory only, see \cite{Schneider:Cui:2005}.
}.
Performing the integral, the new Hamiltonian $\hat{\HAM{}{}}$
written in terms of the {\em transformed} variables $\hat{X}$
will have the form (taking the theorem of interchange-of-variables
in the Hamiltonian as input)
\begin{eqnarray}
\label{Eq::HamiltonianTransformed}
\hat{\HAM{}{}}(\hat{X}) &\stackrel{\rm split \,H}{=}& \HAM{}{}(\hat{X}) + 
\sum_{k\ge 1}^{} \frac{1}{k!}{\cal D}_s^k 
\left( F^* + R \right)
\neanl
&\stackrel{\rm split \,k=1}{=}&
F^{*}(\hat{X})
+\underbrace{R(\hat{X}) + {\cal D}_s^1 F^*}_{= R + \PB{F^*}{s} = 0}
\neanl &&+ 
\sum_{k\ge 2}^{} \frac{1}{k!}{\cal D}_s^k 
F^*
+ \sum_{k\ge 1}^{} \frac{1}{k!}{\cal D}_s^k 
R
\neanl &\stackrel{m=k-1}{=}& 
F^{*}(\hat{X})
+
\sum_{m \ge 1}^{} \frac{1}{(m+1)!}{\cal D}_s^{m} 
\underbrace{\PB{F^*}{s}}_{=-R}
\neanl &&+  
\sum_{k\ge 1}^{} \frac{1}{k!}{\cal D}_s^k 
R
\neanl &=& 
F^{*}(\hat{X})
+
\underbrace{\sum_{k\ge 1}^{} \frac{k}{(k+1)!}{\cal D}_s^k R}_{\bydefd\,K, {\rm ~now~of~order ~\order{\epsilon}{2}}}
\,.
\end{eqnarray}
because $\order{R}{}=\epsilon$ and $\order{s}{}=\epsilon$.
In other words: the new Hamiltonian will be just the old integrable
part (with simply the new variables as arguments) and a higher-order
series involving Lie derivatives of the oscillatory part. Let us call
this new term $K(\hat{X})$.
This last part may contain terms which are integrable and can be absorbed,
for a second transformation, into $F^{*(1)}$ as
\begin{equation}
F^{*(1)}(\hat{X}) \defdby F^*(\hat{X}) + {\rm integrable ~part ~of~} K(\hat{X})\,.
\end{equation}
This may be repeated until the remaining error is only of some required order.
For the reader's convenience, an application of the method to the Duffing-oscillator
to fourth order in the smallness parameter is provided in \cite{Mai:Schneider:Cui:2008}.

\noindent
{\bf A remark on the form of our rest term:}
{\em The non-integrable remaining term that appeared in 
Ref. \cite{Mai:Schneider:Cui:2008}, see their Eq.~(5.33),
has been modified in our computation. The reason is 
the appearance of terms of higher orders in 
$\epsilon$ -- emanating from their definition
of $s$ -- that we are unable to control.}

\noindent
{\bf A further remark on our notation:}
{The coordinates $X_a$ and the Hamiltonians $F^*$ will get an ordering number
-- superscript $(n)$ -- after the $n^{\rm th}$ transformation. The terms $R$
will have a superscript $(1)$ from the beginning on to emphasize that they already
contribute to the first transformation and are at least of the lowest order of the smallness
parameter.}
All (say $n$) performed Lie transformations have to be combined to obtain a connection
of the final $X^{(n)}$ to the initial $X^{(0)}$. This can be obtained
by Eq.~(5.14) of Ref.~\cite{Mai:Schneider:Cui:2008}, where it is stated how to
combine two Lie transformations with two generating functions to get only one,
and which can be extended to multiple transformations.

For the case of spinning compact binaries evolving in circular orbits, the problem
of solving the equations of motion linear in spin will turn out to fulfill our 
requirements to work with Lie transformations to get rid of complicated oscillatory
terms.

\begin{widetext}
\section{Application to a system with two spins}
\label{Sec::LieTrafo_BinaryNLOSO}
\subsection{Involved Hamiltonians}
The point-mass Hamiltonians to second post-Newtonian accuracy \cite{Schafer:1985}
will be given below.
\begin{eqnarray}
 \HAM{PM}{N}
&=&
\frac{p_r^2}{2}+
\frac{\ang^2}{2 r^2}-\frac{1}{r}
\,, \\
 \HAM{PM}{1PN}
&=&
\cInv{2} \left\{
   \frac{{\ang}^4 (3 \eta -1)}{8 r^4}-\frac{{\ang}^2
   (\eta +3)}{2 r^3}+\frac{1}{2 r^2}
+
p_r^2 \left(\frac{(3 \eta -1) L^2}{4 r^2}-\frac{2 \eta +3}{2 r}\right)+\frac{1}{8} (3 \eta
   -1) p_r^4
\right\}
\,, \\
 \HAM{PM}{2PN}
&=&
\cInv{4} \biggl\{
   \frac{\ang^6 (5 (\eta -1) \eta +1)}{16
   r^6}-\frac{\ang^4 \left(3 \eta ^2+20 \eta -5\right)}{8
   r^5}+\frac{\ang^2 (8 \eta +5)}{2 r^4}+\frac{-3 \eta -1}{4 r^3}
\neanl &+& 
 p_r^6\frac{1}{16} (5 (\eta -1) \eta +1)
+p_r^4 \left(\frac{-8 \eta ^2-20 \eta +5}{8 r}+\frac{3 (5 (\eta -1) \eta +1) L^2}{16
   r^2}\right)
\neanl &+& 
 p_r^2 \left(\frac{\left(-4 \eta ^2-20 \eta +5\right) L^2}{4 r^3}+\frac{3 (5
   (\eta -1) \eta +1) L^4}{16 r^4}+\frac{11 \eta +5}{2 r^2}\right)
\biggr\}
\,.
\end{eqnarray}
%
The terms linear in spin through next-to-leading order read \cite{Damour:Jaranowski:Schafer:2008:1}
\begin{eqnarray}
\HAM{LO}{SO} &=&  \frac{\cInv{3}}{4 r^3}
 \left\{ 
 \left( 2 \eta +3 \sqrt{1-4 \eta }+3 \right) {\LSOne} \okay
+\left( 2 \eta -3 \sqrt{1-4 \eta }+3 \right) {\LStwo} \okay
 \right\} \,,
\\
\HAM{NLO}{SO} &=&
\frac{ \cInv{5} }{16 r^4} \Biggl\{
\LSOne
\biggl[
\rel \left(\left(6 \eta ^2+26 \eta -5\right)
   \pSq+6 \eta  (2 \eta +1)
   p_r^2\right)
\neanl && 
+\sqrt{1-4 \eta } \left(\rel
   \left(6 \eta  p_r^2+(16 \eta -5)
   \pSq\right)-8 (2 \eta +5)\right)-8 (4 \eta +5)
\biggr]
\neanl &&
+\LStwo 
\biggl[
\rel \left(\left(6 \eta ^2+26 \eta -5\right)
   \pSq+6 \eta  (2 \eta +1)
   p_r^2\right)
\neanl && 
+\sqrt{1-4 \eta } \left(\rel
   \left((5-16 \eta ) \pSq-6 \eta 
   p_r^2\right)+8 (2 \eta +5)\right)-8 (4 \eta
   +5)
\biggr]
\Biggr\} \okay
\,,
\end{eqnarray} 
\end{widetext}
where $\vct{p}^2 =\scpm{\vnunit}{\vct{p}}^2 + \left(\vnunit \times \vct{p}\right)^2
= p_r^2 + \frac{\Ang^2}{\rel^2}$,
making use of $p_r \defdby \scpm{\vnunit}{\vct{p}}$, 
$\vct{\rel} = \vnunit \, \rel$, and $\vAng \defdby \vct{\rel} \times \vct{p}$.
None of these terms depend on the orbital $\varphi$ {and the magnitude $\Ang$ is unaltered}.
The spin-orbit Hamiltonians can be expressed in terms of
\begin{eqnarray}
\vct{\Sigma} \defdby \vSone + \vStwo  ~~~~ {\rm and} ~~~~
\vct{\Delta} \defdby \vSone - \vStwo \,,
\end{eqnarray}
where one can represent the scalar products $\LSigmaS$ and $\LDeltaS$
in terms of the new canonical variables. 

This gives
\begin{eqnarray}
 \LSigmaS &=& \frac{1}{2} \left( J^2 - \ang^2 - \SpinTot^2 \right) \,, \\
 \LDeltaS &=& 
 \frac{1}{2 \SpinTot^2} \Bigl\{ (S_1-S_2) (S_1+S_2)
   \left(-\Ang^2+J^2-\SpinTot^2\right)
\neanl &-& 
\sin (\phiSpin) A(J,L,S)
\Bigr\}\,,
\end{eqnarray}
having introduced the shorthand
\begin{eqnarray}
A(J,L,S) 
&=& 
 4 L S S_1 S_2 \sin \alpha_{\text{ks}}(J, L, S) \sin \alpha_{12}(S, S_1, S_2) \,,
\neanl &=:& A \,. 
 \end{eqnarray}
For further computations we also introduce the abbreviations
\begin{eqnarray}
G_4(J,L,S_1,S_2)
&\defdby&
 \left(S_1-S_2\right) \left(S_1+S_2\right)
\neanl &&
 \times (J-\ang) (\ang+J)
\,.
\end{eqnarray}
and
\begin{eqnarray}
  A_J	&=& - 4 L^2 S^2 \sin^2 \alpha_{\text{ks}}(J, L, S) \,, \\
  A_S	&=& - 4 S_1^2 S_2^2 \sin^2 \alpha_{12}(S, S_1, S_2) \,.
\end{eqnarray}
It holds $A^2 = A_J A_S$ and
\begin{eqnarray}
  A_J	&=& (J - L - S) (J + L - S) \neanl && \times (J - L + S) (J + L + S) \,, \\
  A_S	&=& (S - S_1 - S_2) (S + S_1 - S_2) \neanl && \times (S - S_1 + S_2) (S + S_1 + S_2) \,.
\end{eqnarray}
The following subsection will deal with circular orbits and the question how
one can evaluate the Poisson brackets imposing this restriction.

\begin{widetext}
\subsection{Evaluating the Poisson Brackets in the Circular Case}
\noindent
For evaluating the Poisson brackets, we find it important to mention
that for some variable 
$X~\in~{\cal M}~=~\{ \varphi, \Ang, \Upsilon,J,\phiSpin,\SpinTot, \alpha_a, S_a \}$
({$\cal M$} is the phase space elements without $p_r$ and $\rel$) and an arbitrary 
function $Z$ of {$\cal M$} it holds
\begin{equation}
\label{Eq::PBCirc}
\PB{X}{Z} \vline_{\stackrel{\rm p_r=0 ~} {~r=\bar{r}(\Ang,...)} } =
\PB{X}
   {Z \vline_{\stackrel{\rm p_r=0 ~} {~r=\bar{r}(\Ang,...)} }} \,,
\end{equation}
which means that we can evaluate $\PB{\sim}{\sim}$ for two
quantities on a general orbit and later impose the condition
of circularity, replacing $\rel$ by its solution
$\bar{\rel}(\Ang,...)$ to $\dot{p}_\rel \stackrel{!}{=}0$,
or we can do it the other way round.
The case of circularity is no primary constraint and works
without artificial/external forces.
The solution to $\dot{p}_r\stackrel{!}{=}0$ can be given as follows:
\begin{eqnarray}
\bar{r}
&=&
L^2-4 \cInv{2}
\neanl &+& 
\cInv{3} \Biggl\{
    \frac{21\left(J^2-\SpinTot^2\right)}{16 L^2}-\frac{21}{16}
   +\epsilon  
   \left(
-\frac{9 \left(
   A
   \sin(\phiSpin)-\left(J^2-\SpinTot^2\right)
   \left(S_1^2-S_2^2\right)\right)}{4 \Ang^2
   \SpinTot^2}-\frac{9 \left(S_1^2-S_2^2\right)}{4
   \SpinTot^2}
   \right)
\neanl &+& 
   \epsilon^2 \left(
\frac{3}{4}-\frac{3 \left(J^2-\SpinTot^2\right)}{4
   \Ang^2}
\right)
   \Biggr\}
+\cInv{4} \left\{
   -\frac{43 \epsilon ^2}{8 L^2}
   -\frac{253}{32 L^2}
   \right\}
+\cInv{5} 
   \Biggl\{
  -\frac{1605}{256 L^2}
  +\frac{1605 \left(J^2-\SpinTot^2\right)}{256 L^4}
\neanl &-& 
   \epsilon 
   {\frac{295}{32}}
   \left(
   \frac{\left(
   A
   \sin(\phiSpin)-\left(J^2-\SpinTot^2\right)
   \left(S_1^2-S_2^2\right)\right)}
   {\Ang^4 \SpinTot^2}
   +\frac{
   \left(S_1^2-S_2^2\right)}
   {\Ang^2 \SpinTot^2}
   \right)
+ \epsilon^2
   {\frac{173}{32}}
   \left(
   \frac{1}{\Ang^2}
   -\frac{\left(J^2-\SpinTot^2\right)}{\Ang^4}
   \right)
\neanl &-& 
   \epsilon^3 
   {\frac{19}{8}}
   \left(
    \frac{\left(
   A
   \sin(\phiSpin)-\left(J^2-\SpinTot^2\right)
   \left(S_1^2-S_2^2\right)\right)}
   {\Ang^4\SpinTot^2}
   +\frac{
   \left(S_1^2-S_2^2\right)}
   {\Ang^2 \SpinTot^2}
   \right)
\neanl &+& 
 \epsilon ^4 \left(
\frac{3}{16 \Ang^2}-\frac{3\left(J^2-\SpinTot^2\right)}{16 \Ang^4}
   \right)
   \Biggr\}
\,. \okay
\end{eqnarray}
As we decide for the second choice in Eq. (\ref{Eq::PBCirc}),
this is inserted into the total Hamiltonian $\HAM{}{}$, which will also generate
terms of order $\order{c}{-3}$ and $\order{c}{-5}$. The Lie transformation terms
and the terms dictating the angular velocity $\Omega_s$ (defined by $\phiSpin = \Omega_s t + \phiSpin{}_0$)
will origin only in those odd powers of $c$.
We next set 
\begin{equation}
\eta \bydefd \frac{1}{4}-\epsilon^2
\end{equation}
where $\epsilon \ll 1$ in order to emphasize that we work in the region of almost equal masses.
We are able to write down the circular-orbit Hamiltonians that contribute to the spin-orbit part
(=those having odd powers on $c$) in terms of $\epsilon$,
\begin{eqnarray}
\label{Eq::H-Circ-LO-SO}
\HAM{LO}{SO,\,circ}  
&=& \cInv{3} \left\{ \frac{1}{L^6}\JLSGSum 
\left(
 \frac{3 \epsilon \left(S_1^2-S_2^2\right)}{4 \SpinTot^2}
-\frac{\epsilon^2}{4}+\frac{7}{16}
\right)
-\frac{3 \epsilon
   A
    \sin ({\phiSpin})}
  {4 L^6 \SpinTot^2}
\right\}
\,, \okay \\
\label{Eq::H-Circ-NLO-SO}
\HAM{NLO}{SO,\,circ}
 &=&
   \frac{\cInv{5} }{L^8}
   \Biggl\{
   \JLSGSum \left(\frac{99 \epsilon 
   \left(S_1^2-S_2^2\right)}{16
   \SpinTot^2}+\frac{3 \epsilon ^4}{16}-\frac{93
   \epsilon ^2}{32}+\frac{975}{256}\right)-\frac{99
   \epsilon 
   A
   \sin ( \phiSpin )}{16 \SpinTot^2}
   \Biggr\}
\,, \okay
\end{eqnarray}
and for the sake of completeness, those coming from the point-mass parts only,
\begin{eqnarray}
 \label{Eq::H-Circ-NPP}
 \HAM{N}  {PM, \,circ} &=& -\frac{1}{2 \Ang^2} \,, \okay \\
 \label{Eq::H-Circ-1PNPP}
 \HAM{1PN}{PM, \,circ} &=& \cInv{2}\frac{4 \epsilon ^2-37}{32 \Ang^4} \,,  \okay\\
 \label{Eq::H-Circ-2PNPP}
 \HAM{2PN}{PM, \,circ} &=& \cInv{4}\frac{-16 \epsilon ^4-104 \epsilon ^2-1269}{256 \Ang^6} \,, \okay
\end{eqnarray}
The expressions (\ref{Eq::H-Circ-LO-SO})-(\ref{Eq::H-Circ-2PNPP}) are {\em not} a Taylor series in $\epsilon$ where higher orders have been neglected.
In fact, higher orders do not exist.

\subsection{Initial Decomposition}
We shall sketch the Lie transformation procedure representatively for the first
order and truncate our procedure to the order $\order{\epsilon}{3}$, because
terms of order $\order{\epsilon}{4}$ are in direct competition with the second-order
Lie transformation.
The first ``perturbing'' Hamiltonian with purely oscillatory character
may be all that comprises terms of $\HAM{SO}{circ}$ with $\sin (\phiSpin)$:
\begin{eqnarray}
F^{*(0)}
&=& \HAM{N}{PN, circ}+\HAM{N}{1PN, circ}+\HAM{N}{2PN, circ}
\neanl &+&
   \frac{ \cInv{3} }{\Ang^6}
   \left\{
\left(\frac{3 (S_1^2-S_2^2) \epsilon }{4
   \SpinTot^2}+\frac{1}{16} \left(7-4 \epsilon
   ^2\right)\right)
   \right\} \left( J^2 - \Ang^2 - \SpinTot^2 \right)
\neanl &+& 
   \frac{\cInv{5}}{\Ang^8}
   \left\{
\left(\frac{99 (S_1^2-S_2^2) \epsilon }{16
   \SpinTot^2}+\frac{3}{256} 
   \left(
   -248 \epsilon^2
   +325\right)
   \right)
   \right\} \left( J^2 - \Ang^2 - \SpinTot^2 \right)
   \,, \okay
      \\
   R^{(1)} &=&
	-\frac{3 
        A }{16 \Ang^8 \SpinTot^2}
	\sin(\phiSpin) 
	\left\{4 \Ang^2 \cInv{3} \epsilon +33\cInv{5} \epsilon \right\}
   \,, \okay
\end{eqnarray}
because, as we compute the Poisson brackets of $\SpinTot$ and $\phiSpin$, we see that
\begin{eqnarray}
\left\{ \SpinTot , F^{*(0)} \right\} &=& 0 \,,
\eanl
\left\{ \phiSpin , F^{*(0)} \right\} &=& \Omega_s^{(0)} \,,
\eanl
\Omega_s^{(0)} & \defdby&
\frac{\cInv{3}}{L^6}
   \left\{ 
   \frac{1}{8} \SpinTot \left(4 \epsilon
   ^2-7\right)-\frac{3 \epsilon 
   G_4(J,\Ang,S_1,S_2)}{2
   \SpinTot^3}
   \right\}
\neanl &+&
\frac{\cInv{5}}{L^8} 
\left\{
-\frac{99 \epsilon 
   G_4(J,\Ang,S_1,S_2)}{8
   \SpinTot^3}-\frac{3}{128} \SpinTot
   \left(
   -248 \epsilon^2
   +325
   \right)
\right\} \okay
\neanl &=& {\rm const.}
\end{eqnarray}
in (applying our sense of rotation) accordance with the equal-mass case ($\epsilon \rightarrow 0$) evolution equation for $\phiSpin$, Eq.~(4.6d) in Ref.~\cite{Tessmer:2009},
where geometrical considerations lead to
\begin{equation}\label{Eq::Geometry}
\Ang/\sin(\alpha_{\rm ks}-\Theta) = \SpinTot/\sin{\Theta}\,,
\end{equation}
see Fig.~\ref{Fig:Geometry}.
The first generating function is defined to be the time integral of
$R^{(1)}$ to 
some time $t$, so we set $\phiSpin = \Omega_S^{(0)} t + \phiSpin{}_{0}$
with some irrelevant $\phiSpin{}_{0}$ and $\SpinTot = \SpinTot{}_0$; 
then we perform the time integral,
and re-express $\Omega_S^{(0)} t + \phiSpin{}_{0} \rightarrow \phiSpin$.
Next, we compute the new Hamilton function according to 
Eq.~\eqref{Eq::HamiltonianTransformed}
and get $F^{*(1)} ( X^{(1)} ) = F^{*(0)} ( X^{(1)}) + {R}^{(2)}( X^{(1)}) $
with
\begin{eqnarray}
 {R}^{(2)}( X^{(1)}) &=&
 \frac{1}{2!} \PB{{R}}{s^{(1)}}
\neanl &+&
 \frac{2}{3!} \PB{\PB{{R}}{s^{(1)}}}{s^{(1)}}
\neanl &+&
 \frac{3}{4!} \PB{\PB{\PB{{R}}{s^{(1)}}}{s^{(1)}}}{s^{(1)}}
\neanl &+&...
\end{eqnarray}
Notice that the R above possesses integrable contents.
These contents are all that remains when one removes the trigonometric functions of $\phiSpin$ in
\begin{eqnarray}
 R^{(2)} &=& \underbrace{\alpha(X^{(1)})}_{\rm integrable} 
+
\sum_{m,n}	
\underbrace{\beta_m(X^{(1)}) \sin ( m \phiSpin)+
\gamma_n(X^{(1)}) \cos ( n \phiSpin)}_{\rm purely ~periodic} \,,
\quad \{m,n\} \in \mathbb{N}.
\end{eqnarray}
As we absorb these terms into $F^{*(1)}$, the new Hamiltonian after the first transformation
will look as what follows after a short break where we compare our calculation to the aligned-spin
case for convenience of the reader.

\begin{figure}[tc]
  \includegraphics[width=\textwidth, angle=-0,scale=0.4]{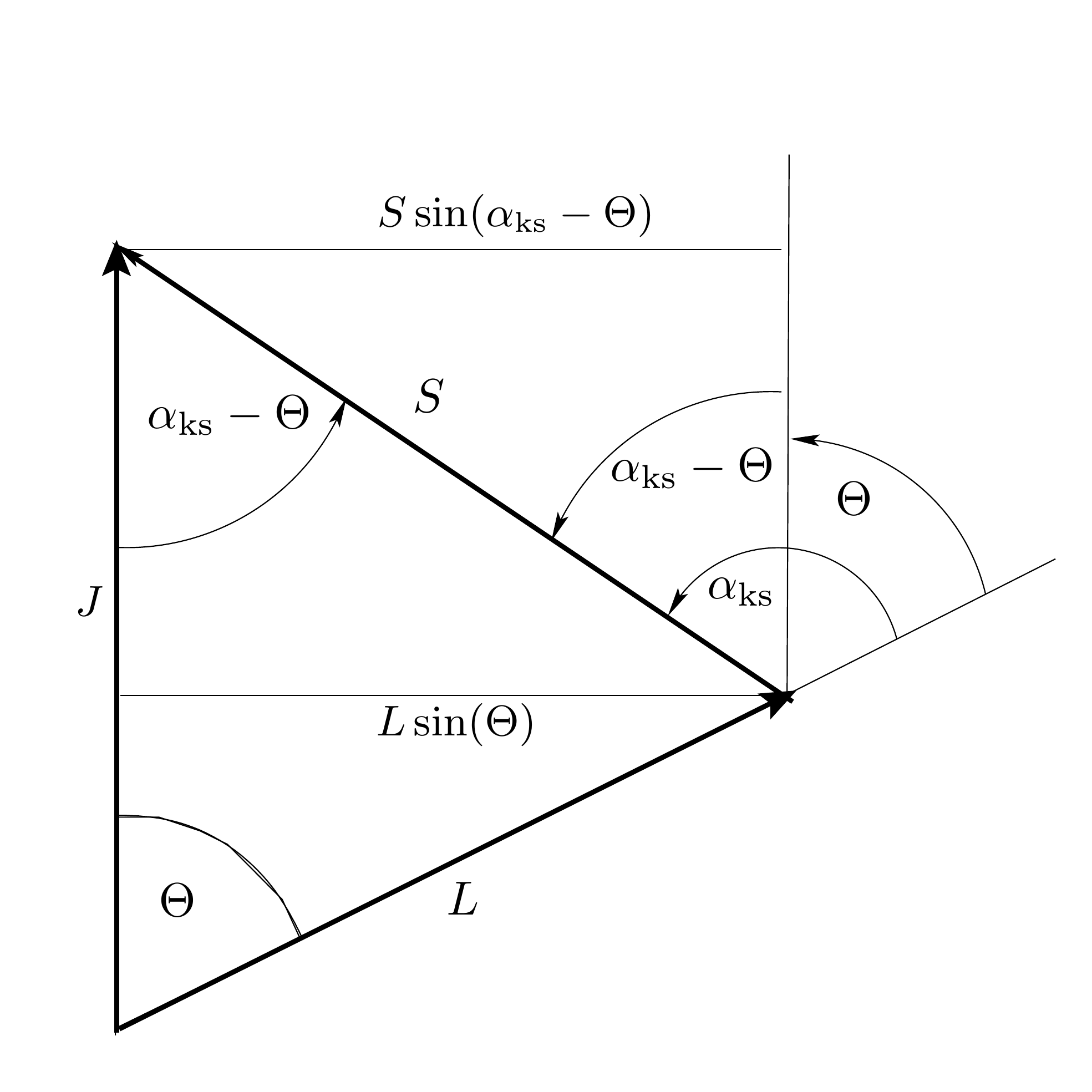}
\caption{The plane spanned by $\vAng$ and $\vct{J}$. Adding auxiliary lines, we
         immediately see that \eqref{Eq::Geometry} holds.\label{Fig:Geometry}}
\end{figure}

\vspace{.5cm}
\subsection{Comparison to the aligned-spin case}
To strengthen the faith in our system of canonical coordinates and conjugated
momenta, and to using the Poisson brackets after imposing circular orbits,
we may do a small check.
In the case of circular orbits at ``almost alignment'', we see that
the spherical coordinates ($\rel, \phi, \theta$) satisfy $r={\rm const}$,
$\Theta \ll 1$, $\theta \approx \frac{\pi}{2}$, and
$\phi = \Upsilon + \varphi + \order{\Theta}{2}$, see Eqs.~(4.27) of
\cite{Konigsdorffer:Gopakumar:2005}. 
As well, we can impose $\SpinTot \approx J - \Ang$ and $\dot{\Theta} \ll 1$,
which can be justified by conservation of $J$ and what triangular relations
for the polygon ($J,\Ang,S_1,S_2$) constrain, for further calculations.

In \cite{Tessmer:Hartung:Schafer:2010}, it has been assumed that $m_1 > m_2$, which is
in contradiction to our assumption. This changes two signs in the function $\eta=\eta(m_1,m_2)$
compared to our computation. We also have to re-scale the spins according to
the rules given therein. Note that the conservation of the dimensionless
$\vct{J}$ is only possible if the individual spins are scaled in the same way $\vAng$ is.

Having done this, we compare our result (TSS) for the spherical phase velocity
to the orbital phase velocity taken from \cite{Tessmer:Hartung:Schafer:2010} (THS),
defining $\Phi$ to be the elapsed total phase through one radial period ${\cal P}_r$,
and see -- through 1PN order linear in spin --
\begin{eqnarray}
 \frac{\Phi}{ {\cal P}_r}_{\rm THS, circ}
 &=& 
   \frac{1}{\Ang^3}
  +\frac{1}{\Ang^6 c^2} 
   \left(
   -\frac{\epsilon ^2 (\Ang+5(S_1+S_2))}{2}
   -{10 \epsilon (S_1-S_2)}
   +\frac{37 \Ang-35(S_1+S_2)}{8}\right)
   \,,  \okay \\
 \dot{\phi}_{\rm TSS, circ}^{\rm re-sc} \approx \left( \dot{\Upsilon} + \dot{\varphi} \right)_{\rm TSS}
 &=&   \frac{1}{\Ang^3}
  +\frac{1}{\Ang^6 c^2} 
   \left(
   -\frac{\epsilon ^2 (\Ang+5(S_1+S_2))}{2}
   -{10 \epsilon (S_1-S_2)}
   +\frac{37 \Ang-35(S_1+S_2)}{8}\right) \,, \okay
\end{eqnarray}
which shows agreement.

\subsection{After the First Transformation}
To omit the superscript ``$(1)$'', the arguments $X_a$ are
we  replace each term by its {transformed version}, $X_a \rightarrow X^{(1)}_a$,
on both sides of the equation:
\begin{eqnarray}
\label{Eq::F*_after_1_trafo}
 F^{*(1)} &=& \left. F^{*(0)} \right|_{X_a \rightarrow X_a^{(1)}}
\neanl &+& 
   \cInv{3}
   \Biggl\{
\FFourJLSg \Biggl(\FFourSaSbSg \left(\frac{243 \epsilon
   ^3 G_4(J,\Ang,S_1,S_2)}{98 \Ang^6
   \SpinTot^{10}}-\frac{45 \epsilon ^2}{56 \Ang^6
   \SpinTot^6}\right)
   +\epsilon ^3 
\frac{54}{49 \Ang^6}
\left[
	 \frac{ S_1^2	}{\SpinTot^8}
	+\frac{ S_2^2	}{\SpinTot^8}
	-\frac{1	}{\SpinTot^6}\right]
   G_4(J,\Ang,S_1,S_2)
\neanl &+& 
   \epsilon ^2
\frac{9}{14 \Ang^6}
   \left[
	 \frac{ 1	}{ \SpinTot^2}
	-\frac{ S_1^2	}{ \SpinTot^4}
	-\frac{ S_2^2	}{ \SpinTot^4}
\right]\Biggr)
+\FFourSaSbSg \Biggl(
\epsilon^3
\frac{54}{49}
\left[
	 \frac{J^2}{ \Ang^6 \SpinTot^8}
	-\frac{1}{ \Ang^6 \SpinTot^6}
	+\frac{1}{ \Ang^4 \SpinTot^8}
\right]
   G_4(J,\Ang,S_1,S_2)
\neanl &+& 
\epsilon^2
\frac{9}{14}
\left[
	 \frac{1}{ \Ang^6 \SpinTot^2}
	-\frac{J^2}{ \Ang^6 \SpinTot^4}
	-\frac{1}{ \Ang^4 \SpinTot^4}
\right]
\Biggr)
\Biggr\}
 +\cInv{5} \Biggl\{
\FFourJLSg \Biggl(\FFourSaSbSg 
\left(
\frac{99873 \epsilon^3 
	 G_4(J,\Ang,S_1,S_2)}{5488 \Ang^8\SpinTot^{10}}
	-\frac{39285 \epsilon ^2}{6272 \Ang^8\SpinTot^6}
\right)
\neanl &+& 
\epsilon ^3
\frac{11097}{1372 \Ang^8 }
\left[
	 \frac{ S_1^2}	{ \SpinTot^8}
	+\frac{ S_2^2}	{ \SpinTot^8}
	-\frac{1}	{ \SpinTot^6}
\right]
   G_4(J,\Ang,S_1,S_2)
+\epsilon ^2
\frac{7857~}{1568 \Ang^8}
\left[
	 \frac{1	}{ \SpinTot^2}
	-\frac{ S_1^2	}{ \SpinTot^4}
	-\frac{ S_2^2	}{ \SpinTot^4}
\right]
\Biggr)
\neanl &+& 
\left.
   \FFourSaSbSg 
\left(
\epsilon^3
\frac{11097}{1372}
\left[
	 \frac{ J^2}{ \Ang^8 \SpinTot^8}
	-\frac{1}{ \Ang^8 \SpinTot^6}
	+\frac{1}{ \Ang^6 \SpinTot^8}
\right]
   G_4(J,\Ang,S_1,S_2)+\epsilon ^2
\frac{7857}{1568}
   \left[
	 \frac{1}{ \Ang^8 \SpinTot^2}
	-\frac{J^2}{ \Ang^8 \SpinTot^4}
	-\frac{1}{ \Ang^6 \SpinTot^4}
   \right] \right)
   \Biggr\} \right|_{X_a \rightarrow X_a^{(1)}}
   \,,
\neanl &&
\\
R^{(2)} &=& \order{\epsilon}{2}\,. \okay
\end{eqnarray}

\subsection{After the Second Transformation}
As one performed a second transformation, the perturbing function $R$
would be (after the split, naturally) shifted to fourth order in $\epsilon$,
and the integrable part also absorbed terms of order $\epsilon^4$,
\begin{eqnarray}
\label{Eq::F2Integrable}
F^{*(2)} \left( X^{(2)} \right)	&=& F^{*(1)}\left( X^{(2)} \right)
   + \order{\epsilon}{4} \,, \\
R^{(3)} \left( X^{(2)} \right)	&=&  \order{\epsilon}{4} \,.
\end{eqnarray}
We will not perform this transformation and stop the calculation here.
All emanating residues are of the form $\sim \sin ( m \phiSpin)$ and 
$\sim \cos(m\phiSpin)$ with $m$ and $n$ as positive integers and having
complicated functions of spin amplitudes as total prefactors, which one
can easily verify for all orders. The residue after the first
transformation is of order $\order{\epsilon}{2}$ -- when talking about
(\ref{Eq::F*_after_1_trafo}), we speak of an integrable system of
{\em first order} in the perturbation parameter $\epsilon$.
Its solution (labeled with a ``bar'') reads

\begin{equation}
\begin{tabular}{lclcrl}
$\bar{{\phi}}_S  (t)$ &=& $\Omega_{S} t$	&+& $ \phiSpin(t=0) \,,$ &
\qquad $\Omega_S := \{ \phi_S^{(1)} , F^{*(1)} \}$ \,, \\
$\bar{{\varphi}} (t)$ &=& $\Omega_\Ang t$	&+& $ \varphi(t=0) \,,$ &
\qquad $\Omega_\Ang := \{ \varphi^{(1)} , F^{*(1)} \}$ \,, \\
$\bar{{\Upsilon}}(t)$ &=& $\Omega_\Upsilon t$	&+& $ \Upsilon(t=0) \,,$ &
\qquad $\Omega_\Upsilon := \{ \Upsilon^{(1)}_{} , F^{*(1)} \}$ \,, \\
$\bar{{S}}     (t)$ &=& $~$			& & $\SpinTot (t=0)\,.$ &
 \end{tabular}
\end{equation}
Because the generator only affects the remaining term, the integrable part
is unaffected, and the circular point mass Hamiltonians always keep their form
as they belong to $F^*$ (up to the fact that the variables get new names).
The reader should note that there are ``Newtonian'' terms in the generating functions,
clearly speaking: terms of order $\order{c}{0}$, which arise because of the fact
that the precession velocity is of the order $\order{c}{-2} \order{\epsilon}{0}$,
and when integrating $R$, these velocities become some part of the denominator.

\vspace{.5cm}
\section{Combining Further Canonical Transformations}
\label{Sec::CombiningLieTrafos}
From Section 11.2.3 of Ref.~\cite{Mai:Schneider:Cui:2008} we know that
combining two Lie transformations with generators $s^{(1)}$ and $s^{(2)}$
will be expressible as performing a single Lie transformation with the generator
\begin{eqnarray}
 s^{(2,1)} &=& (s^{(1)} + s^{(2)}) + \frac{1}{2}\PB{s^{(1)}}{s^{(2)}} 
   + \frac{1}{12} \PB{s^{(1)} - s^{(2)}}{\PB{s^{(1)}}{s^{(2)}}}
\dots
\end{eqnarray}
The above transformation connects the variables $\vct{X}^{(2)}$ (those after the second transformation)
to the initial ones $\vct{X}^{}$ via
\begin{equation}
 \vct{X} = \vct{X}^{(2)} + \frac{1}{2!}\PB{\vct{X}^{(2)}}{ s^{(2,1)} } + \frac{1}{3!}\PB{\PB{\vct{X}^{(2)}}{ s^{(2,1)} }}{ s^{(2,1)} }
\dots
\end{equation}
The transformation may be inverted and the resulting $s^{(2,1)}$ be expressed
entirely in terms of the initial $\vct{X}$. 
We will skip this formula because of reasons of comprehensibility and state that, if
the reader is interested in further transformations due to the reduction of oscillatory
remainder functions, the total generating function of $n$ successive Lie transformations can be obtained from the recursion scheme
\begin{eqnarray}
s^{(1)} &=& s^{(1)} \,, \\
 s^{(2,1)} &=& (s^{(1)} + s^{(2)}) + \frac{1}{2}\PB{s^{(1)}}{s^{(2)}} 
   + \frac{1}{12} \PB{s^{(1)} - s^{(2)}}{\PB{s^{(1)}}{s^{(2)}}} + ...
\,,  \\
\vdots \nonumber \\
s^{(n, n-1)} &=& (s^{(n-1)} + s^{(n)}) + \frac{1}{2}\PB{s^{(n-1)}}{s^{(n)}} 
   + \frac{1}{12} \PB{s^{(n-1)} - s^{(n)}}{\PB{s^{(n-1)}}{s^{(n)}}} +...
\,.
\end{eqnarray}
Structurally, as we would take the full spin-orbit Hamiltonian as input to our
scheme without truncating after third order of the smallness parameter $\epsilon$,
the residues emanating after the n$^{\rm th}$ iteration are going to appear at the
following orders of $\epsilon$:
\begin{table}[!hc]
\begin{tabular}{|c||c|c|c|c|c|c|c|l}
\hline
 step $\#$ 
 & $\epsilon^1$ & $\epsilon^2$ & $\epsilon^3$ & $\epsilon^4$ 
 & $\epsilon^5$ & $\epsilon^6$ & $\epsilon^7$ & $\epsilon^8$ \dots \\
\hline \hline
0 &*& &*& & & & & \\
\hline
1 & &*&*&*&*&*&*&~* \\
\hline
2 & & & &*&*&*&*&~* \\
\hline
3 & & & & & & & &~* \\
\hline
\end{tabular}
\caption{
Position of terms contributing to the oscillatory residuum, 
starting from the {\em untruncated} spin-orbit Hamiltonian,
Eqs.~(\ref{Eq::H-Circ-LO-SO})~and~(\ref{Eq::H-Circ-NLO-SO}).
Step 0 means the initial form. From step 1 onwards,
there are infinitely many terms at higher orders
of $\epsilon$.
}
\end{table}

\section{Some Remarks about higher orders}
In \cite{Schneider:Cui:2005} it was stated that
the Lie series converges if 
there exists a finite number $B$
such that the generating function satisfies
\begin{equation}
\label{Eq::Condition_of_Convergence}
 \left| s \right| <B \,, \left| \frac{\partial^{k_1 + k_2 ... m_1 ... + m_n} s}
{\partial_{x_{i_1}^{k_1}}
\partial_{x_{i_n}^{k_n}}
...
\partial_{y_{j_1}^{m_1}}
...
\partial_{y_{j_n}^{m_n}}}\right|
<B
\end{equation}
For simplicity, we use the perturbation function at linear order in $\epsilon$
and first PN order,
which is of structure
\begin{eqnarray}
 s^{(1)} &=& \frac{3 \epsilon  \left(51 \cInv{2}-112 \Ang^2\right)
   A
   \cos(\phiSpin)}{392 \Ang^2 \SpinTot^3}
+\order{\epsilon}{3}
\end{eqnarray}
However, the quantity A generates more and more terms
at each evaluation of the Poisson bracket which grow in their magnitude.
One can see this fact as well in Eqs. (\ref{Dalpha1}) -- (\ref{Dalpha4}),
where the sines can have values arbitrarily close to zero.
This circumstance violates Eq.~(\ref{Eq::Condition_of_Convergence})
and thus generates  -- in general -- an asymptotic Lie series.
It depends on the system (i.e.: the initial configuration) and on the
mass parameters how many
terms of this asymptotic series can be taken to properly describe
the dynamics of the spin-orbit problem\footnote{A first numerical
insight showed that the range of $\SpinTot(t)$ due to the initial
Hamiltonian depends on the initial value $\phiSpin(t=0)$. Some values
lead to the full range $J-\Ang \le \SpinTot \le S_1+S_2$ while others
lead to low-range oscillations around the initial spin length
$\SpinTot(t=0)$, which may also affect the speed of divergence.
}.

Let us give a numerical representation of the results and their speed of divergence
through a sequence of 3 modified Lie transformations
for an example set
\begin{eqnarray}
 \vct{X}_{\rm num}^{(3)} = \left\{ \SpinTot^{(3)}=1, S_1^{(3)}=1, S_2^{(3)}=\frac{1}{2}, 
 \Ang^{(3)}=10, J^{(3)}=11 \right\}\,.
\end{eqnarray}
Then we obtain a value of
\begin{eqnarray}
F^{*(3)}_{\rm num} &=&
-0.0050
\neanl &+& 
\cInv{2}
10^{-4}
\left\{
-1.15
+0.13 \epsilon ^2
\right\}
\neanl &+& 
\cInv{3}\left\{ 
 0.0000086
+0.0000113 \epsilon
+0.0001276 \epsilon^2
-0.0035799 \epsilon^3
+0.0377074 \epsilon^4
+3.2521845 \epsilon^5
\right\}
\neanl &+& 
\cInv{4}
10^{-6}
\left\{
-4.957 
-0.406 \epsilon ^2
-0.063 \epsilon ^4
\right\}
\neanl &+& 
\cInv{5}
\left\{
 0.00000076
+0.00000093 \epsilon
+0.00000975 \epsilon^2
-0.00026274 \epsilon^3
+0.00259869 \epsilon^4
+0.20891229 \epsilon^5
\right\}
\neanl &+& 
\order{\epsilon}{6} \,.
\end{eqnarray}

\end{widetext}
\section{Conclusions and Outlook}
\label{Sec::Conclusions}
In this article, we have found a reduced variable space
for the treatment of the binary spin-orbit interactions.
This space consists of pairs of compact angle variables and their
conjugate momenta, which are amplitudes of angular momenta defining
the plane in which the specific angle is evolving. \\
Taking all spin-orbit interactions through NLO and 2PN point mass contributions,
we could solve the resulting equations of motion fir the circular-orbit case taking as help the Lie
transformation method to successively get rid of (until now) untreated
oscillatory terms.
It showed off that, by construction, the structure of the point mass Hamiltonians is
not affected by the Lie transformation algorithm as there are no
$\scpm{\vct{S}_a}{\vnunit{}}$ couplings in the spin-orbit terms. \\
\noindent
{\em Outlook:} A numerical comparison of the transformed Hamiltonian
to the original one 
may give some insight about the correctness with respect to the order 
of the perturbation parameter $\epsilon$, regarding a number of configurations.
We also concern resummation techniques in the future that
keep the structure of singular points in the perturbing Hamiltonians but
remarkably reduce the number of the involved terms.

It should be straightforward to extend the results of the present paper
to higher post-Newtonian orders, i.e. to include the NNLO spin-orbit
Hamiltonian. It would be valuable to consider other methods and perturbation
parameters $\epsilon$ to analytically obtain solutions to the equations of motion.
As well, as our phase space generates an asymptotic series for the spin-orbit problem,
a new proposal for canonical variables (where no sines of non-canonical orientation angles,
but polynomials appear in the generator) may mean a future improvement. \\
This can also facilitate an extension to higher orders in spin possible, e.g., to
spin(1)-spin(2) interactions, which actually seem to be impossible to be handled
with the current approach. Furthermore an extension to eccentric orbit by an
expansion around circular-orbit case is envisaged.

\acknowledgments
We wish to thank Professor Manfred Schneider for many inspiring discussions
and suggestions during the originating process of the manuscript.
GS thanks the Erwin Schr\"odinger International Institute for Mathematical
Physics at the Universit\"at Wien for hospitality.
Thanks also go to an anonymous referee for bringing to our attention useful references. 
This work is partly funded by the DFG (Deutsche Forschungsgemeinschaft) through
SFB/TR7 ``Gravitationswellenastronomie,'' STE 2017/1-1, and
the Research Training Group GRK 1523 ``Quanten- und Gravitationsfelder,''
and by the DLR (Deutsches Zentrum f\"ur Luft- und Raumfahrt)  through ``LISA Germany.''

\ifnotprd
\bibliographystyle{utphys}
\fi

\input{LieTrafosAndCanonicAngles_refs_arxiv.tex}
\end{document}

%% file: Macros.tex
\def\vct#1{{\mathbf{#1}}}
\def\defdby{:=} 
\def\bydefd{=:} 

\def\eanl{\\}
\def\neanl{\nonumber\\}

\newcommand{\HAM}[2]{	{\rm H}_{\rm {#1}}^{\rm {#2}}}

\newcommand{\vSone}{ \vct{S}_1 }
\newcommand{\vStwo}{ \vct{S}_2 }
\newcommand{\cInv}[1]{ c^{-#1}}

\newcommand{\pSq}{ \vct{p}^2 }
\newcommand{\LSOne}{ \scpm{\vAng}{\vSone} }
\newcommand{\LStwo}{ \scpm{\vAng}{\vStwo} }
\newcommand{\LDeltaS}{ \scpm{\vAng}{\vct{\Delta}} }
\newcommand{\LSigmaS}{ \scpm{\vAng}{\vct{\Sigma}} }
\newcommand{\SpinTot}{ S }
\newcommand{\phiSpin}{ \phi_S}
\newcommand{\FFourSaSbSg}{ A_{S}^{} {}  }
\newcommand{\FFourJLSg}  { A_{J}^{} {}  }
\newcommand{\JLSGSum}	 { \left(J^2-\Ang^2-\SpinTot^2\right) }

\newcommand{\vnunit}{{\vnxa{12}}}

\newcommand{\vnxa}[1]{{\vct{n}}_{#1}}

\newcommand{\scpm}[2]{(#1 \cdot #2)}

\newcommand{\PB}  [2] { \left\{ #1, #2  \right\}    }

\newcommand{\rel}{r}

\newcommand{\order}[2]{{\cal O}({#1}^{#2})}

\newcommand{\ang}{L}
\newcommand{\vAng}{\vct{L}}
\newcommand{\Ang}{L}

\newcommand{\rotm}{\Lambda}
\newcommand{\avel}{\Omega}

\newcommand{\up}{\vphantom{j}}

%% file: LieTrafosAndCanonicAngles_refs_arxiv.tex
%